\documentclass[12pt]{article}
\usepackage{natbib}

\usepackage[margin=1in]{geometry}
\usepackage[latin1]{inputenc}
\usepackage[T1]{fontenc}
\usepackage[english]{babel}
\usepackage{textcomp} 
\usepackage{graphicx}
\DeclareGraphicsExtensions{.jpg,.mps,.pdf,.png,.gif}
\usepackage[french]{varioref} 
\usepackage[]{amsmath,amssymb,amsfonts}

\usepackage{float}
\usepackage{bbm}
\usepackage{bbold}
\usepackage{amsthm}
\usepackage{gensymb}
\usepackage{verbatim}
\usepackage{float}
\usepackage{setspace}
\usepackage{verbatim}
\usepackage{multirow} 
\usepackage{enumerate}

\usepackage{hyperref}
\newcounter{magicrownumbers}

\usepackage{ifsym}
\usepackage{mathptmx}
\usepackage{helvet}
\usepackage{courier}
\usepackage{color,calc,ulem}  
  \usepackage{pgf}
  \usepackage{graphicx}
  \usepackage{multicol}
  \usepackage{amsmath,amssymb}
  \usepackage{bbm} 
  \usepackage{multicol} 
\usepackage{color, colortbl}
\usepackage{lscape}
  
\usepackage{type1cm}         

\usepackage{makeidx}         
\usepackage{graphicx}        
\usepackage{multicol}        
\usepackage[bottom]{footmisc}

\definecolor{blue1}{rgb}{0,0,0.5451} 
  \definecolor{blue2}{rgb}{0.2,0.4,0.65}
\definecolor{DarkBlue}{rgb}{0,0,0.5451}  
\definecolor{darkred}{rgb}{0.75,0.0,0.0}
\definecolor{darkgreen}{rgb}{0.0,0.6,0.0}
\definecolor{darkblue}{rgb}{0.0,0.0,0.6}
\definecolor{darkcyan}{rgb}{0.0,0.6,0.6}
\definecolor{darkmagenta}{rgb}{0.6,0.0,0.6}
\definecolor{darkyellow}{rgb}{0.6,0.6,0.0}
\definecolor{lightred}{rgb}{1.0,0.9,0.9}
\definecolor{lightgreen}{rgb}{0.9,1.0,0.9}
\definecolor{lightblue}{rgb}{0.9,0.9,1.0}
\definecolor{lightcyan}{rgb}{0.8,1.0,1.0}
\definecolor{lightmagenta}{rgb}{1.0,0.8,1.0}
\definecolor{lightyellow}{rgb}{1.0,1.0,0.8}
\definecolor{paleyellow}{rgb}{1.00,1.0,0.80}
\definecolor{amber}{rgb}{1.0,0.8,0.0}
\definecolor{darkamber}{rgb}{1.0,0.5,0.0} 
\definecolor{Gray}{gray}{0.9}

\definecolor{gray}{cmyk}{0,0.1,0.2,0.3,0.4,0.5,0.6.0.7,0.8,0.9}

\newtheorem*{theorem*}{\normalfont  THEOREM}

\newtheorem*{prop*}{\normalfont PROPOSITION}

\newcommand{\Prob}{\mathbb P}

\newcommand{\Es}{\mathbb E}

\newcommand{\R}{r}
\newcommand{\Set}{\mathcal{F}}
\newcommand{\cone}{\mathcal{C}}

\title{High-dimensional peaks-over-threshold inference}
\author{Rapha\"el de Fondeville \thanks{raphael.de-fondeville@epfl.ch} \: and Anthony C. Davison \thanks{anthony.davison@epfl.ch}\\
{\small\textit{Ecole Polytechnique Fédérale de Lausanne, EPFL-FSB-MATH-STAT,}} \\
{\small\textit{Station 8, 1015 Lausanne, Switzerland}}}

\begin{document}
\doublespacing
\maketitle

\begin{abstract}
Max-stable processes are increasingly widely used for modelling complex extreme events, but existing fitting methods are computationally demanding, limiting applications to a few dozen variables.  $\R$-Pareto processes are mathematically simpler and have the potential advantage of incorporating all relevant extreme events, by generalizing the notion of a univariate exceedance.  In this paper we investigate score matching  for performing high-dimensional peaks over threshold inference, focusing on extreme value processes associated to log-Gaussian random functions and discuss the behaviour of the proposed estimators for regularly-varying distributions with normalized marginals. Their performance is assessed on grids with several hundred locations, simulating from both the true model and from its domain of attraction.  We illustrate the potential and flexibility of our methods by modelling extreme rainfall on a grid with $3600$ locations, based on risks for exceedances over local quantiles and for large spatially accumulated rainfall, and briefly discuss diagnostics of model fit. 
The differences between the two fitted models highlight the importance of the choice of risk and its impact on the dependence structure.
\end{abstract}


\section{Introduction}
Recent contributions in extreme value theory describe models capable of handling spatio-temporal phenomena \citep[e.g.,][]{Kabluchko2009} and provide a flexible framework for modelling rare events, but their complexity makes inference difficult, if not intractable, for high-dimensional data. For instance, the number of terms in the  block maximum likelihood for a Brown--Resnick process grows with dimension like the Bell numbers \citep{Huser2013a}, so less efficient but computationally cheaper methods like composite likelihood \citep{Padoan2010} or the inclusion of partition information \citep{Stephenson2005a} have been advocated.  The first is slow, however, and the second is liable to bias if the partition is incorrect \citep{Wadsworth2015}.  

An attractive alternative to use of block maxima is peaks over threshold analysis, which includes more information by focusing on single extreme events. In the multivariate case, specific definitions of exceedances have been used \citep[e.g.,][]{Ferreira2014,Engelke2012b}, which can be unified within the framework of $\R$-Pareto processes \citep{Dombry2013}. For this approach, a full likelihood is often available in closed form, thus increasing the maximum number of variables that can be jointly modelled from a handful to a few dozen, but non-extreme values may be used, leading to biased estimation. Censored likelihood, proposed in this context by \citet{Wadsworth2014}, is more robust with regard to non-extreme observations, but it involves multivariate normal distribution functions, which can be computationally expensive.  Nevertheless, inference is feasible in $30$ or so dimensions.

Nonparametric alternatives to full likelihood inference developed using the tail dependence coefficient \citep{Davis2009,Davis} or the stable tail dependence function \citep{Einmahl2016} rely on pairwise estimators and allow peaks-over-threshold inference in about a hundred dimensions, but are limited by combinatorial considerations.

Applications of max-stable processes \citep[e.g.,][]{Asadi2015} or Pareto processes \citep{Thibaud2013}  have focused on small regions and have used at most few dozen locations with specific notions of exceedance, but exploitation of much larger gridded datasets from global and regional climatological models along with complex definitions of risk is needed for a better understanding of extreme events and to reduce model uncertainties. The goals of this paper are to highlight the advantages of generalized peaks-over-threshold modelling using $\R$-Pareto processes, to show the feasibility of high-dimensional inference for the Brown--Resnick {model} with hundreds of locations, and to compare the robustness of  different procedures with regard to finite thresholds. We develop an estimation method based on the gradient score \citep{Hyvarinen2005} for a generalized notion of exceedances, for which computational complexity is driven by matrix inversion, similarly to classical Gaussian likelihood inference.  This method focuses on single extreme events and a general notion of exceedance, modelled by Pareto processes, instead of the max-stable approach.

Section~\ref{sec: theory} reviews recent results on regular variation for continuous processes and {generalized} peaks over threshold theory, with a focus on extreme-value processes associated to log-Gaussian random vectors.  In Section~\ref{sec: HD inference}, classical inference schemes are summarised, an efficient parallel algorithm for maximum likelihood is developed, and a faster alternative based on the gradient score \citep{Hyvarinen2005} is considered. Section~\ref{sec: simulations} describes simulations that establish the computational tractability of these procedures and investigate their robustness. In Section~\ref{sec: case study} we apply our methods to estimate Florida extreme rainfall dependence structure for two types of risks using a grid with $3600$ cells.

\section{Modelling exceedances over a high threshold}\label{sec: theory}
\subsection{Univariate model}
The statistical analysis of extremes was first developed for block maxima \citep[Section 5.1]{Gumbel1958}.
This approach is widely used and can give good results, but the reduction of a complex dataset to maxima can lead to significant loss of information \citep{Madsen1997}, so the modelling of exceedances over a threshold is often preferred in applications \citep{Davison1990}. Let $X$ be a random variable with distribution function $F$ satisfying Theorem~3.1.1 in \citet[][Section 3.1 p. 48]{Coles2001}. Then for a large enough threshold $u > 0$,
\begin{equation}\label{eq: gpd}
\mathbb{P}\left( X-u > x \mid X>u\right) \approx H_{(\xi,\sigma)}(x)  = \left\{
\begin{array}{ll}
\left(1+\xi x/\sigma\right)_+^{-1/ \xi}, & \xi \neq 0, \\
\exp\left(- x/\sigma\right),  & \xi = 0,
\end{array}
\right. 
\end{equation}
where $\sigma > 0$ and  $a_+ = \max(a, 0)$.
If the shape parameter $\xi$ is negative, then $X$ must lie in the interval $[0,-\sigma/\xi]$, whereas $X$ can take any positive value with positive or zero $\xi$.
The implication is that the distribution over a high threshold $u$ of any random variable $X$ satisfying conditions for equation~(\ref{eq: gpd}) can be approximated by
\begin{equation}
\label{eq:GPD}
G_{(\xi,\sigma, u)}(x) = 1 - \zeta_u H_{(\xi,\sigma)}(x-u), \quad x > u,
\end{equation}
where $\zeta_u$, the probability that $X$ exceeds the threshold $u$, is determined by $u$.
In its simplest form this model for univariate exceedances applies to independent and identically-distributed variables, but it has been used for time series, non-stationary and spatial data. 


Modelling exceedances can be generalized to a multivariate setting \citep{Rootzen2006} and to continuous processes \citep{Ferreira2014, Dombry2013} within the functional regular variation framework.

\subsection{Functional regular variation}

Let $S$ be a compact metric space, such as $[0,1]^2$ for spatial applications. We write $\Set = C\{S,[0,\infty)\}$ for the Banach space of continuous functions $x: S \rightarrow [0,\infty)$ endowed with the uniform norm $\|x\|_\infty = \sup_{s \in S} |x(s)|$ and $\mathcal{B}(\Xi)$ for the Borel $\sigma$-algebra associated to a metric space $\Xi$.
A measurable closed subset $\cone$ of $\Set$  is called a cone if $tx \in \cone$ for any $x \in \cone$ and $t >0$.
For the study of extremes, the cones $\cone = \{0\}$ or $\cone = \{x \in \Set : \inf_{s \in S} x(s) \leqslant 0\}$ are often excluded from $\Set $ to avoid the appearance of limiting measures with infinite masses at the origin or on the coordinate axes, so let $M_{\Set \setminus \cone}$ denote the  class of Borel measures on $\mathcal{B}(\Set \setminus \cone)$ for any cone $\cone$, and say that a set $A \in \mathcal{B}(\Set \setminus \cone)$ is bounded away from $\cone$ if $d(A,\cone) = \inf_{x\in A, y\in \cone} d(x,y) > 0$.
A sequence of measures $\{\nu_n\}\subset M_{\Set \setminus \cone}$ is said to converge to a limit $\nu\in M_{\Set \setminus \cone}$, written $\nu_n \xrightarrow{\hat{w}} \nu$ \citep{Hult2005}, if
$\lim_{n \rightarrow \infty} \nu_n(A) = \nu(A)$, 
for all $A \in \mathcal{B}(\Set \setminus \cone)$ bounded away from $\cone$ with $\nu(\partial A) = 0$, where $\partial A$ denotes the boundary of $A$.
For equivalent definitions of this so-called $\hat{w}$-convergence, see \citet[Theorem 2.1]{Lindskog2014}.

Regular variation provides a flexible mathematical setting in which to characterize the tail behaviour of random processes in terms of $\hat{w}$-convergence of measures.
A stochastic process $X$ with sample paths in $\Set \setminus \cone$ is regularly varying \citep{Hult2005} if there exists a sequence of positive real numbers $a_1,a_2,\ldots$ with $\lim_{n \rightarrow \infty} a_n = \infty$, and a measure $\nu \in M_{\Set \setminus \cone}$ such that
\begin{equation}\label{eq: rv}
n\Prob\left(a_n^{-1}X \in \cdot\right) \xrightarrow{\hat{w}}  \nu(\cdot), \quad n \rightarrow \infty;
\end{equation}
then we write $X \in {\rm RV}\left(\Set \setminus \cone, a_n, \nu\right)$.
For a normalized processes $X^*$, obtained by standardizing marginals of $X$ to unit Fr\'echet \citep[e.g.,][Section 5]{Coles1991} or unit Pareto \citep{Kluppelberg2008}, for instance, regular variation is equivalent to the convergence of the renormalised pointwise maximum $n^{-1} \max_{i = 1,\dots,n} X_i^*$ of independent replicates of $X^*$ to a non-degenerate process $Z^*$, with unit Fr\'echet margins and exponent measure $\nu^*$ \citep{DeHaan2001}.
The process $Z^*$ is called simple max-stable, and $X^*$ is said to lie in the max-domain of attraction of $Z^*$.

Regular variation also impacts the properties of exceedances over high thresholds.  
For any nonnegative measurable functional $\R : \Set \rightarrow [0, +\infty)$ and stochastic process $\{X(s)\}_{s \in S}$, an $\R$-exceedance is defined to be an event $\{ \R(X) > u_n \}$ where the threshold $u_n$  is such that 
$\Prob \{\R(X) > u_n\} \to 0$ as $n \rightarrow \infty$.  
We further require that $\R$ satisfies a homogeneity property, i.e., there exists $\alpha > 0$ such that $\R(ax) = a^\alpha \R(x)$, for $a> 0$ and $x \in \Set$.
\citet{Dombry2013} called $\R$ a `cost functional' and \cite{Opitz2013} called it a `radial aggregation function', but we prefer the term `risk functional'  because $\R$ determines the type of extreme event whose risk is to be studied.

A natural formulation of subsequent results on $\R$-exceedances uses a pseudo-polar decomposition.  
For a norm $\|\cdot\|_{\rm ang}$ on $\Set$, called the angular norm, and a risk functional $\R$, a pseudo-polar transformation $T$ is a map such that
$$
T: \Set \setminus \cone \rightarrow [0,\infty) \times \mathcal{S}_{\rm ang} \setminus T(\cone), \quad T(x) = \left\{r = \R(x), w = \frac{x}{\|x\|_{\rm ang}}\right\},
$$
where $ \mathcal{S}_{\rm ang} $ is the unit sphere $\{ x \in \Set \setminus \cone : \|x\|_{\rm ang} = 1\}$. If $\R$ is continuous and $T$ is restricted to $\{x \in \Set \setminus \cone : \R(x) >0\}$, then $T$ is a homeomorphism with inverse $T^{-1}(r,w) = r\times w/\R(w)$.

Theorem 2.1 in \cite{Lindskog2014} provides an equivalent pseudo-polar formulation of equation~(\ref{eq: rv}). For any $X \in {\rm RV}\left(\Set \setminus \cone, a_n, \nu\right)$ and any uniformly continuous risk functional $\R$ such that $T(\cone)$ is closed and $\R$ does not vanish $\nu$-almost everywhere, there exist $\beta > 0$ and a measure $\sigma_{\R}$ on $\mathcal{B}(\mathcal{S}_{\rm ang})$ such that
\begin{equation}\label{eq: polar rv}
n\Prob\left\{T^{-1}\left(a_n^{-1}r, w\right) \in \cdot\right\} \xrightarrow{\hat{w}} \nu \circ T^{-1}(\cdot) =  \nu_{\beta} \times \sigma_{\R} (\cdot), \quad n \rightarrow \infty,
\end{equation}
where $\nu_{\beta} [r,\infty) = r^{-\beta}$ and the angular measure $\sigma_{\R} (\cdot)$ equals $\nu\left\{x \in \Set \setminus \cone : \R(x) > 1, \: x/\|x\|_{\rm ang} \in (\cdot) \right\}$.
The converse holds if $\{x \in \Set \setminus \cone : \R(x) = 0\} =  \emptyset$ and $\cone$ is compact \citep[Corollary~4.4]{Lindskog2014}.

The functional $\R(x) = \sup_{s \in S}\{x(s)\}$, used  by \citet{Rootzen2006} in a  multivariate setting and by \citet{Ferreira2014} for continuous processes, implies that realisations of $X(s)$ exceeding the threshold at any location $s \in S$ are labelled extreme, but this functional can only be used in applications where $X(s)$ is observed throughout $S$. Thus it may be preferable to use functions such as $\max_{s\in S'} X(s)$ or $\max_{s\in S'} X(s)/u(s)$, where $S'\subset S$ is a finite set of gauged sites. 
Other suggested risk functionals include $\int_{S} X(s) ds$ for the study of areal rainfall  \citep{Tawn1996}, $\min_{s \in S'} X(s)/u(s)$, or $X(s_0)$ for risks impacting a specific location $s_0$.
Although the choice of risk functional allows a focus on particular types of extreme event, the choice of the angular norm $\|\cdot\|_{\rm ang}$ has no impact and is usually made for convenience.

Finally, for a common angular norm $\|\cdot\|_{\rm ang}$, the angular measures of two risk functionals $\R_1$ and $\R_2$ that are strictly positive $\nu$-almost everywhere  are linked by the expression 
\begin{equation}\label{eq: ang measure link}
\sigma_{\R_1} (dw) = \left\{\frac{\R_1(dw)}{\R_2(dw)}\right\}^\beta\sigma_{\R_2} (dw), \quad dw \in \mathcal{B}(\mathcal{S}_{\rm ang}).
\end{equation}
Equation~(\ref{eq: ang measure link}) is useful when we are interested in $\R_2$-exceedances but inference has been performed based on $\R_1$.
All the previous definitions and results also hold for finite dimensions, i.e., for $I$-dimensional random vectors, by replacing $\hat{w}$-convergence by vague convergence \citep[Section 3.3.5]{Resnick2007} on $M_{\mathbb{R}^I \setminus \cone^I}$, the class of Borel measures on $\mathcal{B}(\mathbb{R}^I \setminus \cone^I)$ endowed with the $\|\cdot\|_\infty$ norm, where $\cone^I$ denotes a cone in $\mathbb{R}^I$ \citep{Opitz2013}

\subsection{$\R$-Pareto processes}
In this section, $\R$ denotes a functional that is nonnegative  and  homogeneous of order $\alpha = 1$, $\Set^+$ denotes the restriction of $\Set$ to nonnegative functions and $\cone$ is the closed cone $\{0\}$.
The $\R$-Pareto processes \citep{Dombry2013} are important  for modeling exceedances, and may be constructed as
\begin{equation}
P = U\frac{Q}{\R(Q)},
\end{equation}
where $U$ is a univariate Pareto random variable with $\Prob(U > r) = 1/r^\beta$  $(r\geq 1)$ and $Q$ is a random process with sample paths in $\mathcal{S}_{\rm ang}^+ = \{x \in \Set^+ \setminus \cone: \|x\|_{\rm ang} = 1\}$ and probability measure $\sigma_{\rm ang}$; then $P$ is called an $\R$-Pareto process with tail index $\beta > 0$ and angular measure $\sigma_{\rm ang}$, and we write $P\sim P_{\beta,\sigma_{\rm ang}}^\R$. 

An important property of this class of processes is threshold-invariance: for all $A \in \mathcal{B}(\Set^+)$ and all $u \geqslant 1$ such that $\Prob\{\R(P) > u\} > 0$,
\begin{equation}\label{eq: pot stability}
\Prob\{u^{-1}P \in A \mid \R(P) > u\} = \mathbb{P}(P \in A).
\end{equation}
Furthermore, for $X\in {\rm RV}\left(\Set^+ \setminus \cone, a_n, \nu\right)$ with index $\beta >0$ and for a risk functional $\R$ that is continuous at the origin and does not vanish $\nu$-almost everywhere, the distribution of the $\R$-exceedances converges weakly to that of a Pareto process, i.e.,
\begin{equation}\label{eq: weak conv Pareto}
\Prob\left\{u^{-1}X \in (\cdot) \mid \R(X) > u\right\} \xrightarrow{w} P_{\beta,\sigma_{\R}}^\R, \quad u \rightarrow \infty,
\end{equation}
with tail index $\beta$ and probability measure $ \sigma_{\R}$ as defined in equation~(\ref{eq: polar rv}) \citep[Theorem~2]{Dombry2013}.
When working with a normalized process $X^*$, the exponent measure $\nu^*$ of the limiting max-stable process $Z^*$ and the measure $\nu_1 \times \sigma_{\R}$ of the Pareto process are equal up to a coordinate transform, as suggested by equation~(\ref{eq: polar rv}).  \citet{Opitz2013} derived these results in a multivariate setting.


\subsection{Extreme value processes associated to log-Gaussian random functions}\label{sec: BR}

We focus on a class of generalized Pareto processes based on log-Gaussian stochastic processes, whose max-stable counterparts are Brown--Resnick processes.
This class is particularly useful, not only for its flexibility but also because it is based on classical Gaussian models widely used in applications; \citet[][p.~84--108]{Chiles2012a} review existing models.

Let $Z$ be a zero-mean Gaussian process with stationary increments, i.e., the semi-variogram $\gamma(s,s')  = \Es[\{Z(s) - Z(s')\}^2]/2$, $(s,s' \in S)$ depends only the difference $s-s'$ \citep[][p.~30]{Chiles2012a}.
If $Z_1, Z_2,\ldots$ are independent copies of a zero-mean Gaussian process with semi-variogram $\gamma$ and $\{U_i : i \in \mathbb{N} \}$ is a Poisson process on $(0, + \infty)$ with intensity $u^{-2}du$, then 
\begin{equation}\label{eq: BR}
M(s) = \max_{ i \in \mathbb{N}} U_i \exp\{Z_i(s) - \gamma(0,s)\}, \quad s \in S,
\end{equation}
is a stationary max-stable Brown--Resnick process with standard Fr\'echet margins, whose distribution depends only on $\gamma$ \citep{Kabluchko2009}.

Let $s_1,\ldots, s_I$ be locations of interest in $S$.
In the rest of the paper, $x$ denotes an element of $\mathbb{R}^I_+$ and $x_i \equiv x(s_i)$ $(i = 1,\dots, I)$ denote its components.
The finite-dimensional exponent measure $\Lambda_{\theta}(\cdot)$ of a simple Brown--Resnick process with $I > 1$ variables is
\begin{equation}\label{eq: exp measure max}
\Lambda_{\theta} (x)= \Es\left[ \max_{i = 1, \dots, I} \left\{ \frac{Z(s_i)  - \gamma(0,s)}{x_i} \right\} \right] = \nu_\theta\left\{A_{\max}(x)\right\},
\end{equation}
where $\nu_\theta(\cdot)$ is the finite-dimensional equivalent of the measure defined in Equation~(\ref{eq: rv}), $\theta$ is an element of the compact set $\Theta$ of the parameters of the semi-variogram $\gamma_\theta$ and $A_{\max}(x) = \left\{y \in \mathbb{R}^I : \max(y_1/x_1, \dots, y_I/x_I) > 1\right\}$.
A closed form for $\Lambda_{\theta} (x)$ is \citep{Huser2013a} 
\begin{equation}\label{eq: brown resnick}
\Lambda_{\theta} (x)= \sum_{i=1}^I \frac{1}{x_i} \Phi\{\eta_i(x), R_i\},
\end{equation}
where $\eta_i$ is the $(I - 1)$-dimensional vector with $j^{th}$ component $\eta_{ij} = \sqrt{\gamma_{i,j}/2} + \log(x_j/x_i)/ \sqrt{2 \gamma_{i,j}} $, $\gamma_{j,k}$ denotes  $\gamma(s_j,s_k)$ $(s_j,s_k \in S)$, and $\Phi( \cdot , R_i)$ is the multivariate normal cumulative distribution function with zero mean and covariance matrix $R_i$ whose $(j,k)$ entry is $(\gamma_{i,j} +\gamma_{i,k} - \gamma_{j,k})/\{2(\gamma_{i,j}\gamma_{i,k} )^{1/2}\}$.

{$\R$-Pareto processes associated to log-Gaussian random functions are closely related to the intensity function $\lambda_{\theta}$ corresponding to the measure $\nu_\theta$, which can be found by taking partial derivatives of $\Lambda_{\theta} (x)$ with respect to $x_1, \dots, x_I$, yielding \citep{Engelke2012b}
\begin{equation} \label{eq: cdfBR Engelke}
\lambda_{\theta}(x) = \frac{| \Sigma_{\theta}|^{-1/2}}{x_1^2x_2 \cdots x_I (2\pi)^{(I-1)/2}} \exp \left( -\frac{1}{2} \widetilde{x}^T \Sigma_{\theta}^{-1} \widetilde{x}\right), \quad x \in \mathbb{R}^I_+,
\end{equation}
where $\widetilde{x} $ is the $(I-1)$-dimensional vector with components $\{\log(x_j/x_1) + \gamma_{j,1} : j = 2, \dots, I \}$ and $\Sigma_{\theta} $ is the $(I-1)\times(I-1)$ matrix with elements $ \{ \gamma_{i,1} + \gamma{j,1} - \gamma_{i,j} \}_{i,j \in \{2, \dots, I\} }$.
\cite{Wadsworth2014} derive an alternative symmetric expression for (\ref{eq: cdfBR Engelke}) which will be useful in Section~\ref{sec: grad score}, but Equation~(\ref{eq: cdfBR Engelke}) is more readily interpreted.
Similar expressions exist for extremal-$t$ processes \citep{Thibaud2013}.}

\section{Inference for $\R$-Pareto processes}\label{sec: HD inference}
\subsection{Generalities}

In this section, $x^1,\ldots, x^N$ are independent replicates of an $I$-dimensional $\R$-Pareto random vector $P$ with tail index $\beta =1$ and
$y^1,\ldots, y^N$, are independent replicates from a regularly-varying $I$-dimensional random vector $Y^*$ with normalized margins.  

As in the univariate setting,  statistical inference based on block maxima and the max-stable framework discards information by focusing on maxima instead of single events.
These models are difficult to fit not only due to the small number of replicates, but also because the likelihood is usually too complex to compute in high dimensions  \citep{Castruccio2014}.
For the Brown--Resnick process, the full likelihood cannot be computed for more than ten variables \citep{Huser2013a}, except in special cases.
When the occurrence times of maxima are available, inference is typically possible up to a few dozen variables \citep{Stephenson2005a}.

Estimation based on threshold exceedances and the Pareto process has the advantages that individual events are used, the likelihood function is usually simpler, and the choice of the risk functional can tailor the definition of an exceedance to the application.  Equation~(\ref{eq: polar rv}) suggests that the choice of risk functional should not affect the estimates, but this is not entirely true, because the threshold cannot be taken arbitrarily high and the events selected depend on the risk functional $\R$, the choice of which enables the detection of mixtures in the extremes and can improve sub-asymptotic behaviour by fitting the model using only those observations closest to the chosen type of extreme event.  For example, we might expect the extremal dependence of  intense local rainfall events to differ from that of heavy large-scale precipitation, even in the same geographical region.

The probability density function of a Pareto process for $\R$-exceedances over the threshold vector $u \in \mathbb{R}_+^I$ can be found by rescaling the intensity function $\lambda_\theta$ by $\nu_{\theta}\{A_{\R}(u)\}$, yielding
\begin{equation} \label{eq: pareto density}
\lambda_{\theta, u}^{\R}(x) = \frac{\lambda_\theta(x)}{\nu_{\theta}\{A_{\R}(u)\}}, \quad x \in A_{\R}(u),
\end{equation}
where $\nu_{\theta}\{A_{\R}(u)\} = \int_{A_{\R}(u)} \lambda_\theta(x) dx$ and $A_{\R}(u)$ is the exceedance region $\left\{x \in \mathbb{R}^I_+ : \R(x/u) > 1 \right\}$.
Equation (\ref{eq: pareto density}) for $\R$-Pareto process inference yields the log-likelihood 
\begin{equation}\label{eq: likelihood}
\ell(\theta; x^1,\ldots, x^N) = \sum_{n = 1}^{N} \mathbb{1}\left\{\R\left(\frac{x^n}{u}\right) > 1\right\}\log\left[ \frac{\lambda_{\theta}(x^n)}{\nu_{\theta}\{A_{\R}(u)\}} \right],
\end{equation}
where division of vectors is component-wise and $\mathbb{1}$ denotes the indicator function. Maximization of $\ell$ gives an estimator $\widehat{\theta}_{\R}(x^1,\ldots, x^N)$ that is consistent, asymptotically normal and efficient.

Numerical evaluation of the $I$-dimensional integral $\nu_{\theta}\{A_{\R}(u)\}$ is generally intractable for high $I$, though it simplifies for some special risk functionals, such as  $\R(x) = \max_{i = 1,\dots,I} x_i$, for which the integral is a sum of multivariate probability functions; see Equation~(\ref{eq: brown resnick}).
Similarly, \citet{Coles1991} pointed out that $\nu_{\theta}\{A_{\R}(u)\}$ is constant and independent of $\theta$ when the risk functional is $\R(x) = I^{-1} \sum_{i = 1,\dots,I} x_i$; \citet{Engelke2012b} called the resulting quantity (\ref{eq: likelihood}) the spectral likelihood. 

In practice observations cannot be assumed to be exactly Pareto distributed; it is usually more plausible that they lie in the domain of attraction of some extremal process.
As a consequence of Theorem~3.1 in \citet{DeHaan1993}, asymptotic properties of $\widehat{\theta}_{\R}(x^1,\ldots, x^N)$ hold for $\widehat{\theta}_{\R}(y^1,\ldots, y^N)$ as  $N \rightarrow \infty$ and $u \rightarrow \infty$ with the number of exceedances $N_u = o(N)$; see  Section~\ref{sec: grad score}.
However, the threshold $u$ is finite and thus low components of $y^i \in A_{\R}(u)$ may lead to biased estimation.  
As it is due to model mis-specification, this bias is unavoidable, and moreover, it grows with $I$, so these methods can perform poorly, especially if the extremal dependence is weak, as it is then more likely that at least one component of $x^i$ will be small \citep{Engelke2012b,Thibaud2013,Huser}.  
The bias can be reduced by a form of censored likelihood proposed in the multivariate setting by \citet{Joe.Smith.Weissman:1992}, and used for the Brown--Resnick model by \citet{Wadsworth2014}, and for the extremal-$t$ process by \citet{Thibaud2013}.
This method works well in practice but typically requires the computation of multivariate normal and $t$ probabilities, which can be challenging in realistic cases if standard code is used.  Some relatively modest changes to the code to perform quasi-Monte Carlo maximum likelihood estimation with hundreds of locations are described in Section~\ref{sec: censored likelihood}.

For spatio-temporal applications, inference for $\R$-Pareto processes must be performed using data from thousands of locations, and in Section~\ref{sec: grad score} we discuss an approach that applies to a wide range of risk functionals, is computationally fast and statistically efficient, and is robust with regard to finite thresholds.

\subsection{Efficient censored likelihood inference}\label{sec: censored likelihood}
\subsubsection{Definition and properties}
Censored likelihood estimation for extreme value process associated to log-Gaussian random functions was developed by \citet{Wadsworth2014} and is based on equation~(\ref{eq: likelihood}) with $\max_{i = 1,\dots,I}\{x_i/u_i\}$ as risk functional and where any component lying below the threshold vector $(u_1,\dots,u_I) > 0 $ is treated as censored.
This estimator has increased variance but reduced bias compared to the spectral estimator. 
For the Brown--Resnick process, the censored likelihood density function, in \citet{Engelke2012b}'s notation, is
\begin{equation} \label{eq: brown resnick cens}
\lambda_{\theta, u}^{\text{cens}}(x) =   \frac{1}{\nu_{\theta}\{A_{\max}(u)\}}\frac{1}{x_1^2x_2 \cdots x_k} \phi_{k-1}(\widetilde{x}_{2:k}; \Sigma_{2:k}) \Phi_{I-k}\{\mu_{\text{cens}}(x_{1:k}),\Sigma_{\text{cens}}(x_{1:k})\}, \quad x \in A_{\max}(u),
\end{equation}
where $A_{\max}(u) =   \{x \in \mathbb{R}^I : \max_{i = 1, \dots, I} (x_i/u_i) > 1 \}$, $k$ components exceed their thresholds, $\widetilde{x}_{2:k}$ and $\Sigma_{2:k}$ are subsets of the variables $\widetilde{x}$ and $\Sigma_\theta$ in equation~(\ref{eq: cdfBR Engelke}), and $\phi_{k-1}$ and $\Phi_{I - k}$ are the multivariate Gaussian density and distribution functions. The mean and covariance matrix for $\Phi_{I-k}$ are
\begin{eqnarray*}
\mu_{\text{cens}}(x_{1:k}) &=&   \{\log(u_j/x_1) + \gamma_{j,1} \}_{j = k+1, \dots, I} - \Sigma_{(k+1):I, 2:k}\Sigma_{2:k,2:k}^{-1}\widetilde{x}_{2:k}, \\
\Sigma_{\text{cens}}(x_{1:k}) &=&  \Sigma_{(k+1):I, (k+1):I} - \Sigma_{(k+1):I, 2:k}\Sigma_{2:k,2:k}^{-1} \Sigma_{ 2:k,(k+1):I}.
\end{eqnarray*}
\cite{Wadsworth2014} derived similar expressions based on equation~(\ref{eq: cdfBR Wadsworth}).
The estimator
\begin{equation}\label{eq: censored estimator}
\widehat{\theta}_{\text{cens}}(y^1,\ldots, y^N) = \text{arg} \max_{\theta \in \Theta} \sum_{n = 1, \dots, N} \mathbb{1}\left\{\max_{i = 1,\dots,I}\left(\frac{y^n_i}{u_i}\right) > 1\right\}\log \lambda_{\theta, u}^{\text{cens}}(\mathrm{y}^n),
\end{equation}
 is also consistent and asymptotically normal as $u \rightarrow \infty$, $N \rightarrow \infty$, $N_u \rightarrow \infty$ with $N_u = o(N)$.
For finite thresholds,  $\widehat{\theta}_{\text{cens}}$ has been found to be more robust with regard to low components \citep{Engelke2012b,Huser}, but it is awkward due to the potentially large number of multivariate normal integrals involved,  thus far limiting its application to $I\lesssim 30$ \citep{Wadsworth2014,Thibaud.etal:2016}.

A useful alternative is composite likelihood inference \citep{Padoan2010,Varin.Reid.Firth:2011} based on subsets of observations of sizes smaller than $I$, which trades off a gain in computational efficiency against a loss of statistical efficiency.  The number of possible subsets increases very rapidly with $I$, and their selection can be vexed, though some statistical efficiency can be retrieved by  taking higher-dimensional subsets.   \citet{Castruccio2014}  found higher-order composite likelihoods  to be more robust than spectral likelihood, but in realistic cases they are limited to fairly small dimensions. Even with $I=9$ they required days of computation.  
\subsubsection{Quasi-Monte Carlo maximum likelihood}\label{sec:QMC}

When maximizing the right-hand side of equation~(\ref{eq: censored estimator}), the normalizing constant $\nu_{\theta}\{A_{\max}(u)\}$, described in equation~(\ref{eq: exp measure max}), and the multivariate normal distribution functions require the computation of multidimensional integrals.
Theorem~7 of \citet{Geyer1994} suggests that we approximate $\widehat{\theta}_{\text{cens}}$ by maximizing 
\begin{equation}\label{eq: monte carlo log like}
\ell^p_{\text{cens}}(\theta) = \sum_{m = 1}^{n} \mathbb{1}\left\{\max\left(\frac{x^m}{u}\right) > 1\right\} \left[ \log \left\{ \frac{\phi_{t-1}(\widetilde{x}_{2:t}; \Sigma_{2:t})}{(x_1^m)^2x^m_2 \cdots x^m_t} \right\} +  \log\frac{\Phi^p_{I-t}\{\mu_{\text{cens}}(x^m_{1:t}),\Sigma_{\text{cens}}(x^m_{1:t})\}}{\Lambda_{\theta}^p (u) } \right],
\end{equation}
where $\Phi^p_{I-t}$ and $\Lambda^p_\theta$ are Monte Carlo estimates of the corresponding integrals based on  $p$ simulated samples, yielding a maximizer $\widehat{\theta}^p_{\text{cens}}$ that converges almost surely to $\widehat{\theta}_{\text{cens}}$  
as $p \rightarrow \infty$.

Classical Monte Carlo estimation for multivariate integrals yields a probabilistic error bound that is $O(\omega p^{-1/2})$, where $\omega = \omega(\phi)$ is the square root of the variance of the integrand $\phi$.
Quasi-Monte Carlo methods can achieve higher rates of convergence and thus improve computational efficiency while preserving the consistency of $\widehat{\theta}^p_{\text{cens}}$.
For estimation of multivariate normal distribution functions, \citet[][Section 4.2.2]{Genz2009} advocate the use of randomly-shifted deterministic lattice rules, which can achieve a convergence rate of order $O(p^{-2 + \epsilon})$ for some $\epsilon > 0$.
Lattice rules rely on regular sampling of the hypercube $[0,1]^I$, taking  
\begin{equation}\label{eq: lattice rule}
\mathrm{z}_q = |2 \times  \overline{( qv + \Delta)} - 1|, \quad q = 1, \dots, p,
\end{equation}
where $\overline{(\mathrm{z})}$ denotes the component-wise fractional part of $\mathrm{z} \in \mathbb{R}^I$, $p$ is a prime number of samples in the hypercube $[0,1]^I$, $v \in \{1, \dots, p\}^I$ is a carefully-chosen generating vector and $\Delta \in [0,1]^I$ is a uniform random shift.
Fast construction rules exist to find an optimal $v$ for given numbers of dimensions $I$ and samples $p$ \citep{Nuyens2006}.
The existence of generating vectors achieving a nearly optimal convergence rate, with integration error independent of the dimension, has been proved and methods for their construction exist \citep{Dick2010}.

Our implementation of this approach applied to equation~(\ref{eq: censored estimator}) and coupled with parallel computing is tractable for $I$ of the order of a few hundred; see Appendix \ref{app: censored likelihood} for details.

\subsection{ Score matching}\label{sec: grad score}

Classical likelihood inference methods require either evaluation or simplification of the scaling constant $\nu_{\theta}\{A_{\R}(u)\}$, whose complexity increases with the number of dimensions.
Hence  we seek alternatives that do not require its computation.

Let $\mathcal{A}$ be a sample space such as $\mathbb{R}_+^I$, and let $\mathcal{P}$ be a convex class of probability measures on $\mathcal{A}$.
A proper scoring rule \citep{Gneiting2007b} is a functional $\delta: \mathcal{P}\times \mathcal{A} \rightarrow \mathbb{R}$ such that
\begin{equation}\label{eq: properness}
\int_\mathcal{A} \delta(g,x)g(x)dx \geqslant \int_\mathcal{A} \delta(h,x)g(x)dx, \quad h,g \in \mathcal{P}.
\end{equation}
The scoring rule is said to be strictly proper if equality in (\ref{eq: properness}) holds only when $g =h $.
A proper scoring rule is a consistent estimator of a divergence measure between two distributions \citep{Thorarinsdottir2013} and can be used for inference.
For a risk functional $\R$, the estimator
\begin{equation}\label{eq: maximum score estimator}
\widehat{\theta}_{\delta, u}^\R(x^1,\ldots, x^N) = \text{arg} \max_{\theta \in \Theta} \sum_{n = 1}^N \mathbb{1}\left\{\R\left(\frac{x^n}{u}\right) > 1\right\} \delta(\lambda_{\theta, u}^{\R}, x^n),
\end{equation}
where $x^1,\ldots, x^N$ were defined at the beginning of Section~\ref{sec: HD inference}, is a consistent and asymptotically normal estimator under suitable regularity conditions \citep[Theorem~4.1]{Dawid2014}.
As a consequence of \citet[Propositions~3.1, 3.2]{DeHaan1993}, these asymptotic properties can be generalized to samples from a regularly-varying random vector with normalized marginals; see Appendix \ref{app: mda score normality}.

\begin{prop*}
Let $1 \leqslant N_u \leqslant N$. Let $y^1,\ldots, y^N$ be independent replicates of a regularly-varying random vector $Y^*$ with normalized marginals and limiting measure $\nu_{\theta_0}$ and let $\delta$ be a strictly proper scoring rule satisfying the conditions of Theorem~4.1 of \citet{Dawid2014}. If $N \rightarrow \infty$ and $N_u \rightarrow \infty$ such that $N_u = o(N)$, then
$$
\sqrt{N_u}\left\{\widehat{\theta}_{\delta, N/N_u}^\R\left(y^1,\ldots, y^N\right) - \theta_0\right\} \rightarrow \mathcal{N}\left\{0, K^{-1}J(K^{-1})^T\right\}
$$
in distribution, where 
\begin{equation}
J = \Es_P \left\{\frac{\partial\delta}{\partial \theta}(\theta_0)\frac{\partial\delta}{\partial \theta}(\theta_0)^T\right\}, \quad 
K= \Es_P\left\{\frac{\partial^2\delta}{\partial \theta^2}(\theta_0) \right\}.
\end{equation}
\end{prop*}

Estimates of the Godambe information matrix  $G = \left\{K^{-1}J(K^{-1})^T\right\}^{-1}$  can be used for inference, and the scoring-rule ratio statistic
$$
W^\delta = 2\left\{\frac{\partial\delta}{\partial \theta}\left(\theta_0\right) - \frac{\partial\delta}{\partial \theta}\left(\widehat{\theta}_{\delta, n/k_u}^\R\right) \right\},
$$
 properly calibrated, can be used to compare models \citep[Section~4.1]{Dawid2014}.

The log-likelihood function is a proper scoring rule associated to the Kullback--Leibler divergence.  Although efficient, it is not robust, which is problematic for fitting asymptotic models like Pareto processes, and the normalizing coefficient $\nu_{\theta}\{A_{\R}(u)\}$ is obtainable only in special cases.
The gradient score \citep{Hyvarinen2005} uses the derivative $\nabla_{x} \log g$, and so does not require computation of scaling constants such as $\nu_{\theta}\{A_{\R}(u)\}$.
\citet{Hyvarinen2007} adapted this scoring rule for strictly positive variables, and we propose to extend it to any domain of the form $A_{\R}(u) = \{x \in \mathbb{R}_+^I : \R(x / u) > 1\}$, using the divergence measure
\begin{equation}\label{eq: gradient divergence}
\int_{A_{\R}(u)} \| \nabla_{x} \log  g(x) \otimes {w}(x) - \nabla_{x} \log  h(x) \otimes {w}(x) \|^2_2 \:g(x)dx,
\end{equation}
where $g$ and $h$ are multivariate density functions differentiable on { $A_{\R}(u) \setminus \partial A_{\R}(u)$, where $\partial A$ denotes the boundary of $A$}, $\nabla_x$ is the gradient operator, ${w}: A_{\R}(u) \rightarrow \mathbb{R}_+^I$ is a positive weight function,  and $\otimes$ denotes the Hadamard product.
If ${w}(\cdot)$ is differentiable on $A_{\R}(u)$, and if for every $i \in \{1, \dots, I\}$, we have
\begin{equation} \label{eq: consitence grad score}
\lim_{x_i \rightarrow a_i(x_1, \dots, x_{i-1}, x_{i+1}, \dots , x_I)} w_i(x)^2 \frac{\partial \log h(x)}{\partial x_i} g(x) - \lim_{x_i \rightarrow b_i(x_1, \dots, x_{i-1}, x_{i+1}, \dots , x_I)} w_i(x)^2 \frac{\partial \log h(x)}{\partial x_i} g(x) = 0,
\end{equation}
where $a_i(x_1, \dots, x_{i-1}, x_{i+1}, \dots , x_I)$ and $b_i(x_1, \dots, x_{i-1}, x_{i+1}, \dots , x_I)$ are respectively the lower and upper bounds of the variable $x_i$ on $A_{\R}(u)$ for fixed $(x_1, \dots, x_{i-1}, x_{i+1}, \dots , x_I)$, then the scoring rule
\begin{equation}\label{eq: grad score new}
\delta_{{w}}(h,x) = \sum_{i =1}^I \left(2w_i(x)\frac{\partial w_i(x)}{\partial x_i} \frac{\partial \log h(x)}{\partial x_i} + w_i(x)^2 \left[ \frac{\partial^2 \log h(x)}{\partial x_i^2} + \frac{1}{2} \left\{\frac{\partial \log h(x)}{\partial x_i} \right\}^2 \right] \right), \quad x \in A_{\R}(u),
\end{equation}
is strictly proper, as is easily seen by modification of \citet{Hyvarinen2007}.  The gradient score for a Pareto process satisfies the regularity conditions of Theorem~4.1 in \citet{Dawid2014}, so the resulting estimator $\widehat{\theta}_w$ is asymptotically normal.  

Two possible weight functions for inference on the Pareto process are
\begin{equation}\label{eq: grad score weights}
\left.
\begin{array}{ll}
w^1_i(x) = & x_i\left[1 - e^{-\R(x/u) - 1} \right], \\
w^2_i(x) = & \left[1 - e^{-3\frac{x_i - u_i}{u_i}} \right]\left[1 - e^{-\R(x/u) - 1} \right],
\end{array}\right\}
\quad i \in \{1, \dots, I\}, 
\end{equation}
where $\R$ is a risk functional differentiable on $\mathbb{R}_+^I$ and the threshold vector $u$ lies in $\mathbb{R}^I_+$.
The weights ${w}^1$ are derived from \citet{Hyvarinen2007}, whereas ${w}^2$ is designed to approximate the effect of censoring by down-weighting components of $x^i$ near the threshold.
These weighting functions are particularly well suited for extremes: a vector $x \in A_{\R}(u)$ is penalized if $\R(x / u)$ is close to $1$, and low components of $x$ induce low weights for the associated partial derivatives.
For these reasons, inference using $\delta_{w}$ with the weighting functions in equation~(\ref{eq: grad score weights}) can be expected to be more robust to low components than is the spectral log-likelihood.
The estimator $\widehat\theta_w$ can be much cheaper to compute than $\widehat\theta_{\text{cens}}$ and can be obtained for any risk functional differentiable on $\mathbb{R}_+^I$.
The gradient score can be applied to any extremal model with a multivariate density function {whose logarithm} is twice differentiable away from the boundaries of its support, and if these display discontinuities on this support then the weighting function ${w}$, chosen such that~\eqref{eq: consitence grad score} is fulfilled, ensures the existence and the consistency of the score.
Expressions for scores for the Brown--Resnick model can be found in Appendix~\ref{app: BR score}, and the performances of these inference procedures are compared in Section~\ref{sec: simulations}.


\section{Simulation study}\label{sec: simulations}

\subsection{Exact simulation}\label{sec: simul Pareto}
The inference procedures and simulation algorithms described below have been wrapped in an R package, \verb+mvPot+ available on \verb+CRAN+.

We first  illustrate the feasibility of high-dimensional inference by simulating generalized Pareto processes associated to log-normal random functions at $I$ locations.
Details of the algorithm can be found in Appendix~\ref{app: pareto sims}.

\begin{figure}[!t]
\begin{center}
\includegraphics[scale = 0.45]{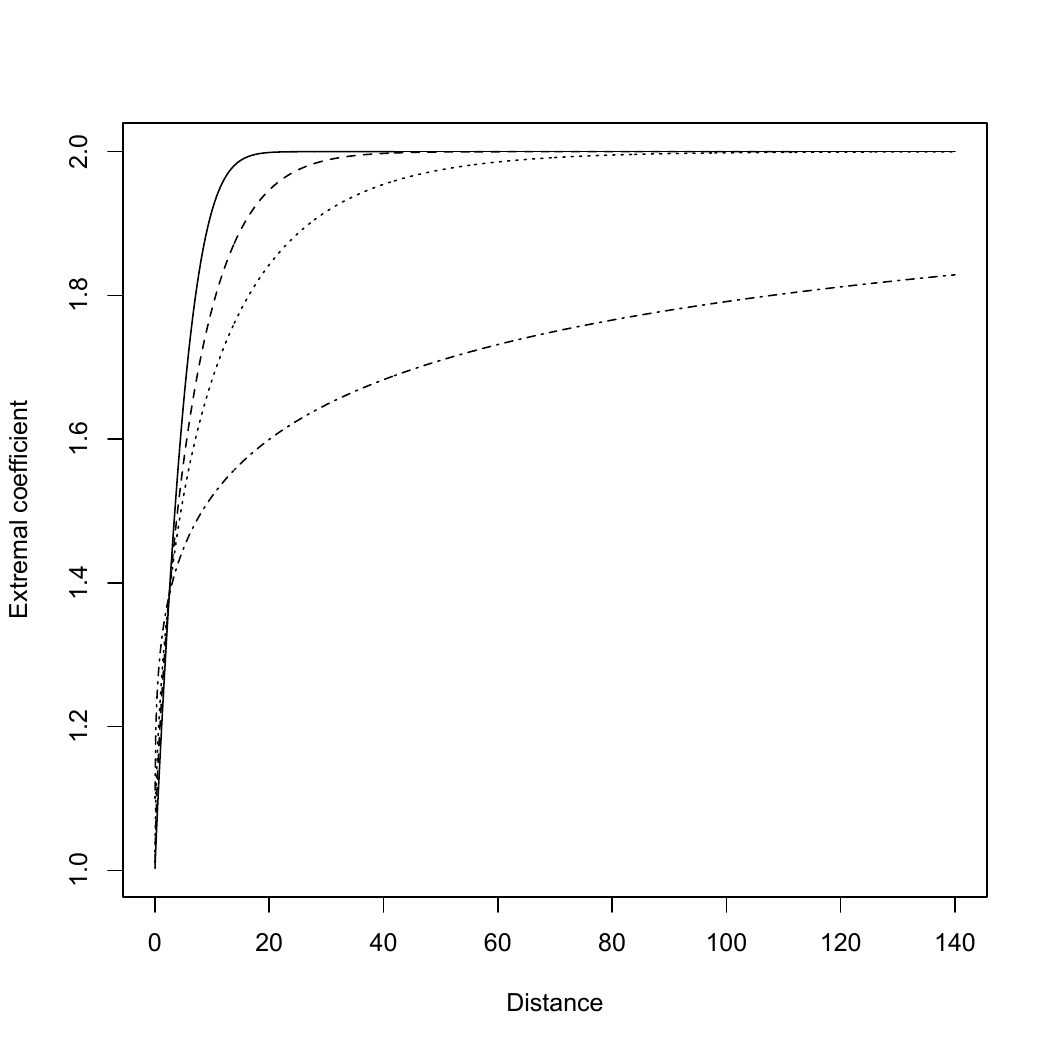} 
\end{center}
\caption{Pairwise extremal coefficient for a Brown--Resnick process  with semi-variogram $\gamma(s,s') = \left(\|s - s'\|/\tau\right)^\kappa$ as a function of distance for $\kappa = 1.8$ (solid), $\kappa = 1.3$ (dashes), $\kappa = 1$ (dots), $\kappa = 0.5$ (dot-dash) and $\tau = 2.5$. The extremal dependence is perfect for an extremal coefficient of $1$ and independence is reached when it equals $2$.}
\label{fig: extremal models}
\end{figure}

We use an isotropic power semi-variogram, $\gamma(s,s') =\left(\|s - s'\|/\tau\right)^\kappa/2$, shape parameters $\kappa= 0.5,1,1.3$, and scale parameter $\tau = 2.5$. In spatial extremes, it is common to compare models by plotting the extremal coefficient \citep{Schlather2003} against distance between locations, as in Figure~\ref{fig: extremal models}. The extremal coefficient measures the strength of dependence, has a lower bound equal to $1$, which is achieved in case of perfect dependence, and an upper bound $2$ corresponding to independence. For this simulation, dependence models with $\kappa > 1.3$  could not be tested because the Pareto process drifts below the smallest representable number and thus rounding produces exact zeros, which are incompatible with the Brown--Resnick model. For each simulation, $N=10,000$ Pareto processes were simulated on regular $10 \times 10$, $20 \times 10$ and $20 \times 15$ grids. The grid size was restricted to a maximum of $300$ locations for ease of comparison with the second simulation study. For the gradient score, we use $\R(x) = \sum_{i = 1}^I x(s_i)$. The threshold $u$ is taken equal to the empirical $0.99$ quantile of $\R(x^1),\ldots, \R(x^N)$, giving $N_u=100$. For censored likelihood inference, we use the approach described in Appendix \ref{app: parallel censored likelihood} with $\bar{p} = 10$. One hundred replicates are used in each case. 

Table~\ref{tab: mvn est Pareto} gives the relative root mean square error for estimation based on censored log-likelihood and the gradient score with weights $w^1$ and $w^2$, relative to that based on the spectral log-likelihood.  For all the methods and parameter combinations, bias is negligible and performance is mainly driven by the variance.
As expected, efficiency is lower than $100\%$ because when simulating and fitting from the true model, the spectral likelihood performs best.  The gradient score and the censored likelihood estimators deteriorate as the extremal dependence weakens and the number of low components in the simulated vectors increases.
The gradient score outperforms the censored likelihood except when censoring is low, i.e., when $\kappa = 0.5$.
The performance of the censored likelihood estimators deteriorates when the dimensionality increases, suggesting that the gradient score will be preferable in high dimensions.   These results, however, are not realistic since the data are simulated from the fitted model, whereas in practice the model is used as a high-threshold approximation to the data distribution.

\begin{table}[!t]
\begin{center}
\begin{tabular}{c}
\begin{tabular}{c c c c c}
 Grid size & $\kappa = 0.5$ & $\kappa = 1$ & $\kappa = 1.3$ \\ \hline
$10 \times 10$  & $\textbf{53.3}/44.8/42.4$  & $10.3/\textbf{30.8}/12.4$ & $4.7/\textbf{36.5}/12.3$ \\ 
$20 \times 10$  & $\textbf{66.8}/48.9/49.0$ & $10.1/\textbf{23.8}/14.1$ & $5.4/\textbf{32.6}/12.4$ \\ 
$20 \times 15$ & $\textbf{66.9}/44.0/43.9$ & $10.6/\textbf{28.8}/16.9$ & $4.1/\textbf{23.4}/9.3$ \\ 
\end{tabular} \\ Shape $\kappa$\\ \\
\begin{tabular}{c c c c c}
 Grid size & $\kappa = 0.5$ & $\kappa = 1$ & $\kappa = 1.3$ \\ \hline
$10 \times 10$  & $52.4/\textbf{54.9}/54.2$  & $18.8/\textbf{57.9}/40.1$ & $10.1/\textbf{58.0}/36.1$ \\ 
$20 \times 10$  & $40.6/\textbf{77.6}/76.2$ & $16.6/\textbf{69.1}/61.2$ & $9.2/\textbf{64.2}/37.2$ \\ 
$20 \times 15$ & $37.9/\textbf{65.6}/67.0$ & $16.5/\textbf{77.6}/64.9$ & $7.1/\textbf{58.2}/29.2$ \\ 
\end{tabular} \\ Scale $\tau$\\
\end{tabular}
\end{center}
\caption{Relative root mean square error (\%) for comparison of estimates based on censored log-likelihood (left) and the gradient score with weights $w_1$ (middle) and $w_2$ (right) relative to those based on the spectral log-likelihood, for the parameters $\kappa$ and $\tau$. Efficiency of $100\%$ corresponds to the performance of the, optimal, maximum spectral log-likelihood estimator, and smaller values show less efficient estimators. Inference is performed using the top 1\% of $10000$ simulated Pareto processes with semi-variogram $\gamma(s,s') =\left(\|s - s'\|/\tau\right)^\kappa/2$. The scale parameter  is $\tau = 2.5$ and grids are regular of sizes $10 \times 10$, $20 \times 10$ and $20 \times 15$ on $[0,100]^2$.} 
\label{tab: mvn est Pareto}
\end{table}

The optimization of the spectral likelihood and gradient score functions takes only a dozen seconds even for the finest grid. The same random starting point is used for each optimization to ensure fair comparison.  Estimation using the censored approach takes several minutes and slows greatly as the dimension increases; see Appendix~\ref{app: computation time}.

\subsection{Domain of attraction}\label{sec: simul BR}

As in practice the asymptotic regime is never reached, we now compare the robustness of each inference procedure for finite thresholds. The Brown--Resnick process belongs to its own max-domain of attraction, so its peaks-over-threshold distribution converges to a generalized Pareto process with log-Gaussian random function. We repeat the simulation study of Section~\ref{sec: simul Pareto} with $10,000$ Brown--Resnick processes and the same parameter values, adding $\kappa = 1.8$. Simulation of the max-stable processes uses the algorithm of \citet{Dombry2016} and is computationally expensive, so we restricted the simulation to $300$ variables. It takes around $3$ hours using $16$ cores to generate $N=10,000$ samples on the finest grid.

\begin{table}
\begin{center}
\begin{tabular}{c}
\begin{tabular}{c c c c c}
 Grid size & $\kappa = 0.5$ & $\kappa = 1$ & $\kappa = 1.3$ & $\kappa = 1.8$ \\ \hline
$10 \times 10$  & $\textbf{153.8}/111.3/80.6$  & $\textbf{472.9}/183.1/107.9$ & $\textbf{196.1}/169.5/105.3$ &  NC\\ 
$20 \times 10$  & $\textbf{171.7}/121.8/95.4$ & $\textbf{413.4}/149.6/113.9$ & $\textbf{308.9}/181.2/136.8$ &  $144.4/\textbf{167.8}/121.8$ \\ 
$20 \times 15$ & $\textbf{142.4}/119.4/99.4$ & $\textbf{369.2}/133.3/109.6$ & $\textbf{313.7}/170.1/139.5$ &  $163.2/\textbf{173.1}/136.6$ \\ 
\end{tabular} \\ Shape $\kappa$\\ \\
\begin{tabular}{c c c c c}
 Grid size & $\kappa = 0.5$ & $\kappa = 1$ & $\kappa = 1.3$ & $\kappa = 1.8$ \\ \hline
$10 \times 10$  & $106.7/\textbf{126.5}/115.6$  & $\textbf{262.49}/38.2/34.6$ & $109.0/231.4/\textbf{451.5}$ &  NC \\ 
$20 \times 10$  & $105.3/\textbf{133.3}/119.2$ & $\textbf{205.7}/94.2/79.7$ & $\textbf{314.8}/65.7/53.2$ &  $104.5/\textbf{335.5}/261.2$ \\ 
$20 \times 15$ & $103.8/\textbf{138.1}/125.9$ & $\textbf{173.4}/101.9/89.7$ & $\textbf{289.5}/91.5/45.8$ &  $102.8/\textbf{211.1}/144.3$ \\ 
\end{tabular} \\ Scale $\tau$\\
\end{tabular}
\end{center}
\caption{Relative root mean square error (\%) for the censored log-likelihood (left) and the gradient score with weights $w_1$ (middle) and $w_2$ (right) relative to those based on the spectral log-likelihood for the parameters $\kappa$ and $\tau$. An efficiency of $100\%$ corresponds to the performance of the maximum spectral log-likelihood estimator, and larger values show more efficient estimators. Inference is based on the top 1\% of $10000$ simulated Brown--Resnick processes with semi-variogram $\gamma(s,s') =  \left(\|s - s'\|/\tau\right)^\kappa/2$. In each case the scale parameter equals $\tau=2.5$ and grids are regular of sizes $10 \times 10$, $20 \times 10$ and $20 \times 15$. ``NC'' means that optimization does not converge.} 
\label{tab: br est efficiency}
\end{table}

Table~\ref{tab: br est efficiency} shows the results.
As expected when the model is misspecified, the root relative mean square error is mainly driven by bias, which increases with the shape $\kappa$ and the dimension $I$.
Spectral likelihood estimation is least robust overall, and for this reason it is outperformed by both other methods.
For $\kappa = 0.5$, the three methods show fairly similar performance, with the censored likelihood better capturing the shape parameter, whereas the gradient score does better for the scale.  The moderate extremal dependence cases, with $\kappa = 1$ and 1.3, are dominated by the censored likelihood, whereas for the weak extremal dependence, $\kappa = 1.8$, the gradient score performs best, because too much information is lost by censoring. For the 100-point grid, the optimization procedures do not converge when the extremal dependence is too weak.  Comparison of the weighting functions $w^1$ and $w^2$ reveals that the choice of the weighting function $w$ affects the robustness of the gradient score. Further simulations, not shown in this paper, show that $w$ tailored to specific types of misspecification can produce very robust estimates.  Computation times are similar to those in Section~\ref{sec: simul Pareto}. 

Quantile-quantile plots show that the score-matching estimators are very close to normally distributed, but censored likelihood estimates can deviate somewhat from normality due to the quasi-Monte Carlo approximation; this can be remedied by increasing the value of $p$.

To summarise: for weak extremal dependence, the three types of estimator are roughly equivalent. For moderate extremal dependence, we recommend using the censored likelihood if the number of variables permits ($I \lesssim 500$ with our computational capabilities), though if extremal independence is reached at far distances and the grid is dense, the gradient score is a very good substitute.  For gridded applications with fine resolution, the gradient score appears to be the best choice for its robustness and because it does not suffer from dimensionality limitations.

\section{Extreme rainfall over Florida}\label{sec: case study}

\subsection{General}

\begin{figure}
\begin{center}
\includegraphics[scale = 0.45]{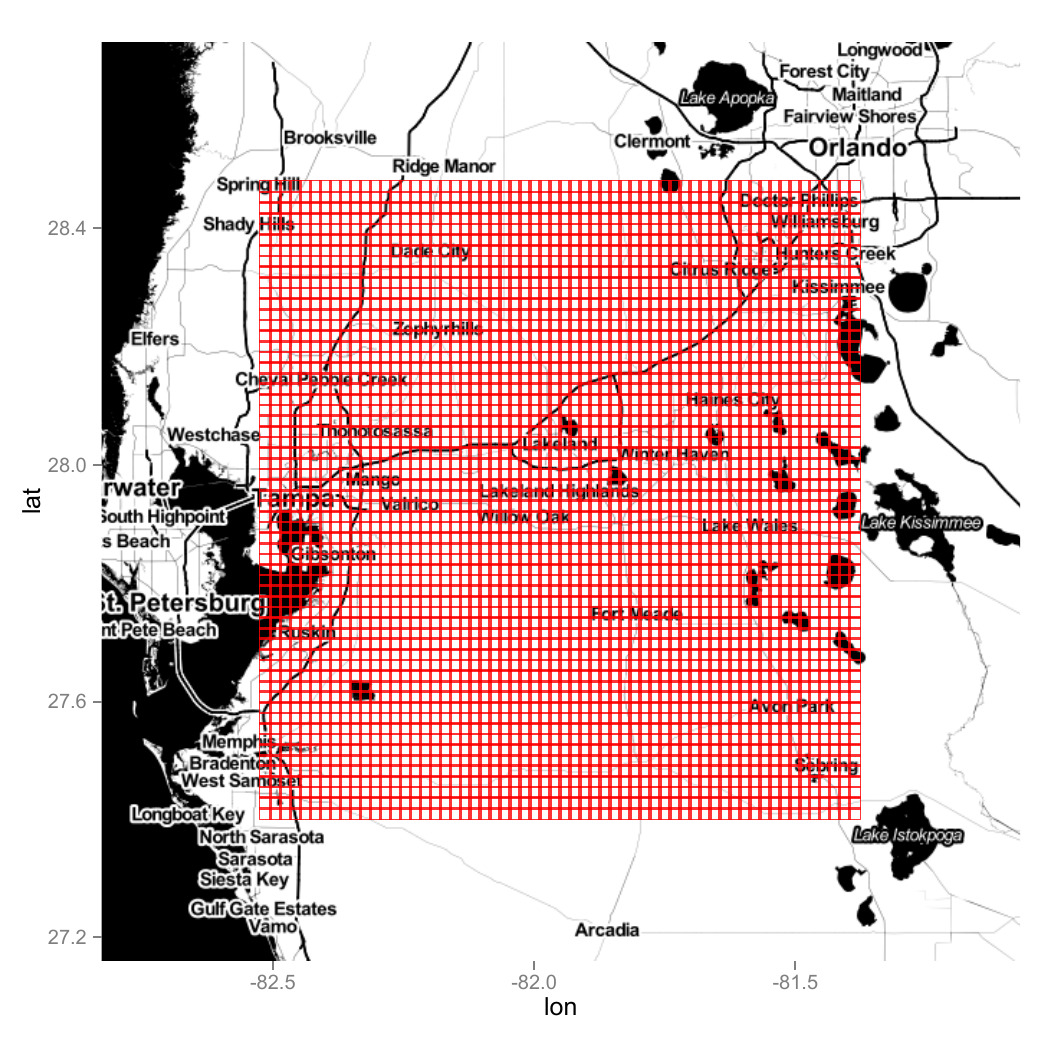} 
\end{center}
\caption{Radar rainfall measurement grid ($2\text{km} \times 2\text{km}$) over East Florida.}
\label{fig: florida}
\end{figure}

We fit a $\R$-Pareto process based on the Brown--Resnick model to radar measurements of rainfall taken every $15$ minutes during the wet season, June--September, from $1999$ to $2004$ on a regular 2~km grid in a 120~km$\times$120~km region of  east Florida; see Figure~\ref{fig: florida}.  There are 3,600 spatial observations in each radar image, and  $58,560$ images in all.  The  region was chosen to repeat the application of \citet{Buhl2015}, but in a spatial setting only; a spatio-temporal model is outside the scope of the present paper.
\citet{Buhl2015} analysed daily maxima for 10~km$\times$10~km squares, but we use non-aggregated data to fit a non-separable parametric model for spatial extremal dependence, using single extreme events instead of daily maxima.  

The marginal distributions for each grid cell were first locally transformed to unit Pareto using their empirical distribution function. For general application, where we wish to extrapolate the distribution above observed intensities, a model for the marginal distributions of exceedances is needed, but since our goal here is to illustrate the feasibility of dependence model estimation on dense grids, we treat marginal modelling as outside the scope of this study.


\subsection{Multivariate extremal dependence model}\label{sec: model fitting}

The spatial model of \citet{Buhl2015} is fully separable, i.e., it is a sum of two separate semi-variograms. This has the advantage that inference for each direction can be performed separately, but it cannot capture any anisotropy that does not follow the axis of the grid, i.e., is not in the South-North or East-West directions.  Furthermore their pairwise likelihood approach focuses on short-distance pairs, and so might mis-estimate dependence at longer distances.
To better capture possible anisotropy, we use the non-separable semi-variogram model
\begin{equation}\label{eq: model}
\gamma(s_i,s_j) = \left\| \frac{\Omega (s_i - s_j)}{\tau} \right\|^\kappa,  \quad s_i,s_j \in [0,120]^2, \quad i,j \in \{1, \dots ,3600\},\quad 0 < \kappa \leqslant 2,\tau>0,
\end{equation}
and anisotropy matrix
\begin{equation}\label{eq: anisotropic matrix}
\Omega = 
\left[\begin{array}{cc}\cos \eta & -\sin \eta \\ a \sin \eta & a \cos \eta\end{array} \right], \quad \eta \in \left(-\frac{\pi}{2}; \frac{\pi}{2}\right], \quad a \geqslant 1.
\end{equation}
The semi-variogram $\gamma$ achieves asymptotic extremal independence as the distance between sites tends to infinity, i.e., the pairwise extremal index $\theta \rightarrow 2$ as $\| s - s'\| \rightarrow  \infty$.

To apply the peaks-over-threshold methodology, we must define exceedances by choosing risk functionals. 
We focus on two types of extremes: local very intense rainfall at any point of the region, and high cumulative rainfall over the whole region, both of which can severely damage infrastructure.
We therefore take the risk functionals
\begin{equation}
\R_{\max}(X^\ast) = \left[ \sum_{i = 1}^I \left\{X^\ast(s_i)\right\}^{20}\right]^{1/20}, \quad
\R_{\text{sum}}(X^\ast) = \left[ \sum_{i = 1}^I \left\{X^\ast(s_i)\right\}^{\xi_0}\right]^{1/ \xi_0} .
\end{equation}
The function $\R_{\text{max}}$ is a differentiable approximation to $\max_{i = 1,\dots,I} X(s_i)$, which cannot be used with the gradient score because of its non-differentiability. Censored likelihood is computationally out of reach with so many locations. 
Directly summing normalized observations $X^*$ makes no physical sense, so we use a modified $\R_{\text{sum}}$, using $\xi_0 = 0.114$, chosen as the mean of independent local estimates of a generalized Pareto distribution; this can be seen as a transformation back to the original data scale.
The function $\R_{\text{sum}}$ selects extreme events with large spatial extent.

We fitted univariate generalized Pareto distributions to $\R_{\text{sum}}(x_m^\ast)$ and $\R_{\max}(x_m^\ast)$ ($m =1, \dots, 58560$) with increasing thresholds.
The estimated shape parameters are stable around the $99.9$ percentile, which we used for event selection, giving $59$ exceedances; 2 events were found to be extreme relative to both risk functionals. Here we merely illustrate the feasibility of high-dimensional inference, so we treat them as independent, but in practice temporal declustering should be considered.

Optimization of the gradient score with the ${w}^1$ weighting function on a $16$-core cluster took from $1$ to $6$ hours, depending on the initial point.
Different initial points must be considered because of the possibility of local maxima.
Results are shown in Table~\ref{tab: model estimates}, where
standard deviations are obtained using a jackknife procedure with $20$ blocks.   Both the estimated bias  and variance are fairly low. {For  $\R_{\text{sum}}(x_m^\ast)$,} we obtain a model similar to that of \citet{Buhl2015}.

The estimated parameters differ appreciably for the two risk functionals, suggesting the presence of a mixture of types of extreme events.
The structure for $\R_{\max}$ is consistent with the database, in which the most intense events tend to be spatially concentrated. 
Our model suggests higher dependence for middle distances than was found by \citet{Buhl2015}, but they note that their model underestimates dependence, especially for high quantiles.
The estimated smoothness parameters are very close.
For $\R_{\text{sum}}$, the estimated parameters shows strong extremal dependence even at long distances, corresponding to exceedances of cumulated rainfall with large spatial cover.
Depending on the risk functional, the model represents either local rainfall, using $\R_{\max}$, or events with wide coverage, using $\R_{\text{sum}}$.
Anisotropy was introduced as in \citet{Buhl2015}, but as $\widehat a\approx 1$, it does not seem necessary.

\begin{table}
\begin{center}
\begin{tabular}{ c   c  c  c  c }
Risk functional & $\kappa$ & $\tau$ & $\eta$ & $a$ \\ \hline
$\R_{\text{sum}}$ & $0.814~(0.036)$ &$25.63~(4.70)$  & $-0.009~(0.458)$ & $1.059~(0.031)$ \\ 
$\R_{\max}$& $0.955~(0.048)$ &$3.54~(0.67)$  & $ -0.316~(0.410)$ & $0.94~(0.029)$ \\ 
\end{tabular}
\end{center}
\caption{Parameter estimates (standard errors) for a Brown--Resnick process with the semi-variogram  $\gamma(s,s') = \left\{\|\Omega(s - s')\|/\tau\right\}^\kappa$ obtained by maximization of the gradient score for events corresponding to $60$ highest exceedances of the risk functionals $\R_{\text{sum}}$ and $\R_{\max}$ for the  Florida radar rainfall data. Standard errors are obtained using a jackknife  with $20$ blocks.} 
\label{tab: model estimates}
\end{table}

\subsection{Model checking and simulation}

\begin{figure}
\begin{center}
\begin{tabular}{cc}

 \raisebox{-.5\height}{\includegraphics[width = 0.45\textwidth]{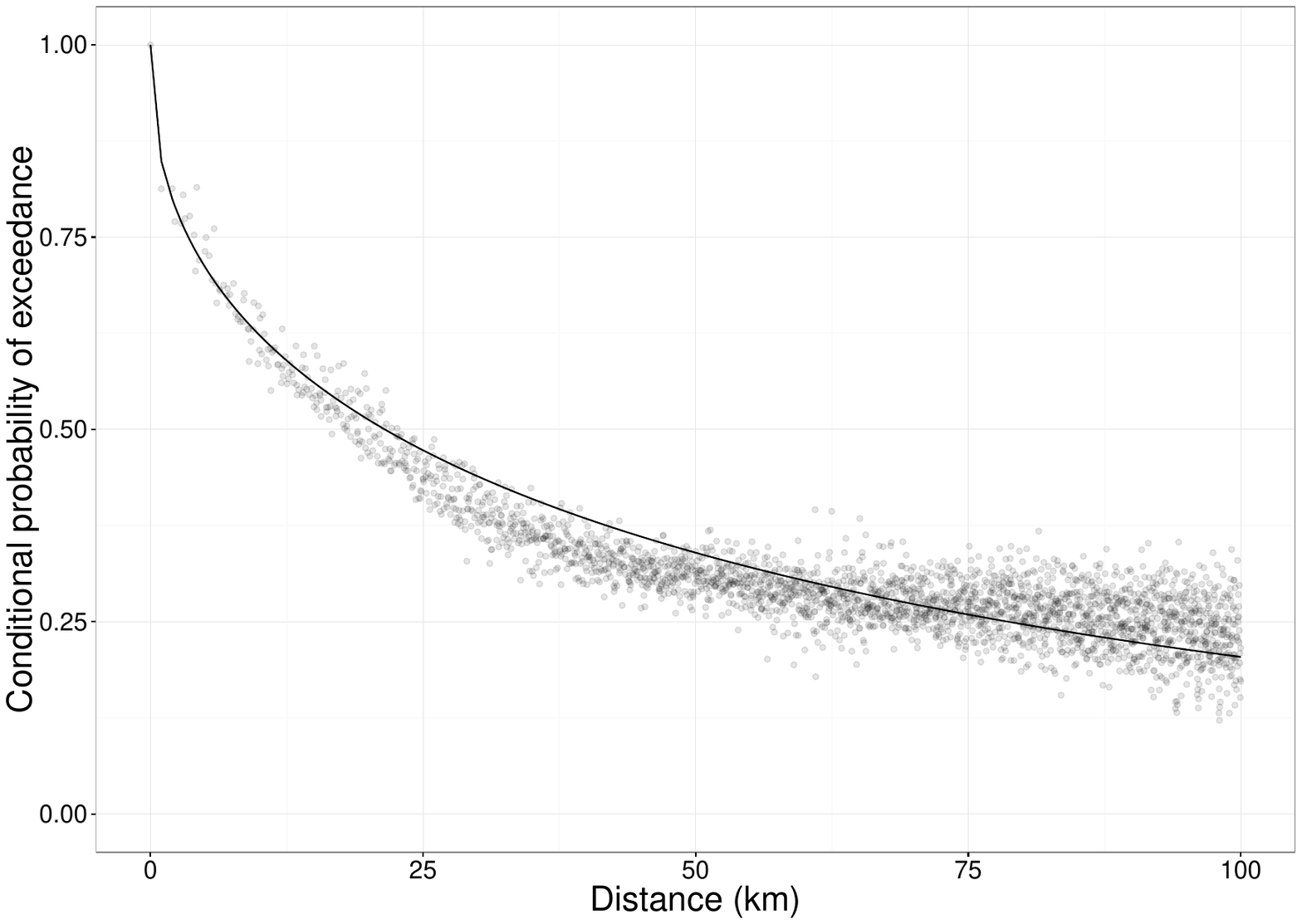}} &  \raisebox{-.5\height}{ \includegraphics[width = 0.45\textwidth]{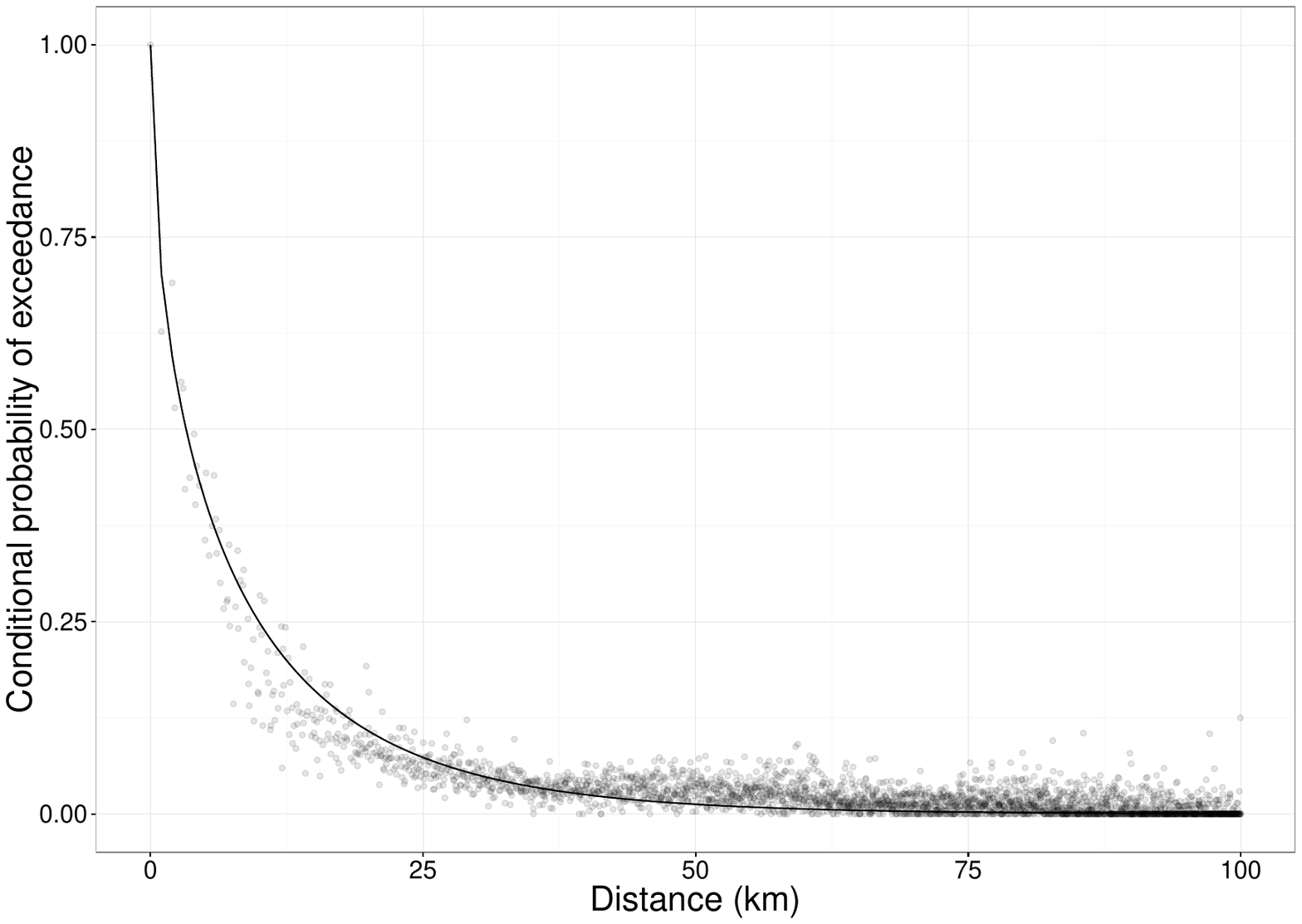} }\\
 ~ \\
 $\R_{\text{sum}}$ & $\R_{\text{max}}$
\end{tabular}
\end{center}
\caption{Estimated conditional probability of exceedance $\pi_{ij} $ for the risk functional $\R_{\text{sum}}$ (left) and $\R_{\text{max}}$ (right) depending on the distance separating locations $s_i$ and $s_j$, $i,j = 1,\dots,3600$. The solid black line represents the model fitted using gradient score estimation.}
\label{fig: model checking}
\end{figure}

For model checking, we propose to use the conditional probability of exceedances, 
\begin{equation}\label{eq: cond prob}
\pi_{ij} = \Pr\left[X^\ast(s_j) > u_j \mid \{X^\ast(s_i ) > u_i\} \cap \{\R(X^\ast/ u)>1\} \right] =  2 \left\{1 - \Phi\left(\sqrt{\frac{\gamma_{ij}}{2}}\right)\right\},
\end{equation}
where $\gamma_{i,j}$ is the semi-variogram for sites $s_i$ and $s_j$  ($i,j = 1,\dots,3600$), as defined in~(\ref{eq: brown resnick}).
A natural estimator for $\pi_{ij}$ is
\begin{equation}\label{eq: cond prob estimator}
\widehat{\pi}_{ij} = \frac{\sum_{n=1}^N  \mathbb{1}\left[\left\{\R\left({x^{\ast n}}/{u}\right) > 1\right\} \cap \left\{ x^{\ast n}_i > u_i \right\}\cap \left\{x^{\ast n}_j > u_j\right\}\right]}{\sum_{n = 1}^N  \mathbb{1}\left[\left\{\R\left({x^{\ast n}}/{u}\right) > 1 \right\}  \cap \left\{ x^{\ast n}_i > u_i\right\}\right]},
\end{equation}
whose asymptotic behaviour can easily be adapted from \citet{Davis2009}.
For both risk functionals, the fitted model, represented by the solid black lines in Figure~\ref{fig: model checking}, follows the cloud of estimated conditional exceedance probabilities reasonably well and captures the general trend, but fails to represent some some local variation, perhaps owing to a lack of flexibility of the power model; a more complex dependence model might be considered.

\begin{figure}
\begin{center}
\begin{tabular}{ccc}
& $\R_{\text{sum}}$ & $\R_{\text{max}}$\\ \hline
\rotatebox[origin=c]{90}{Observations} & \raisebox{-.5\height}{\includegraphics[scale = 0.25]{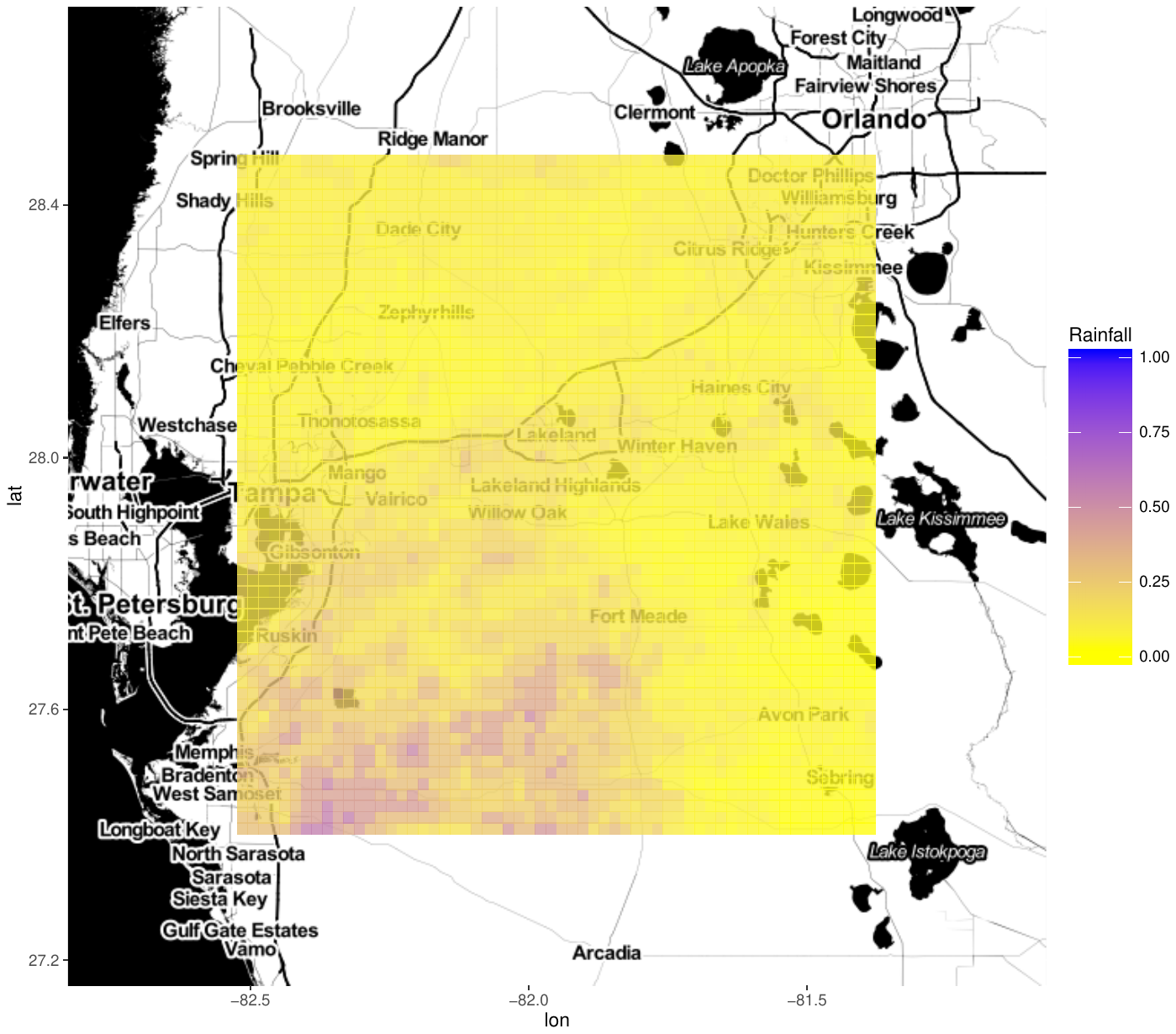}} &  \raisebox{-.5\height}{ \includegraphics[scale = 0.25]{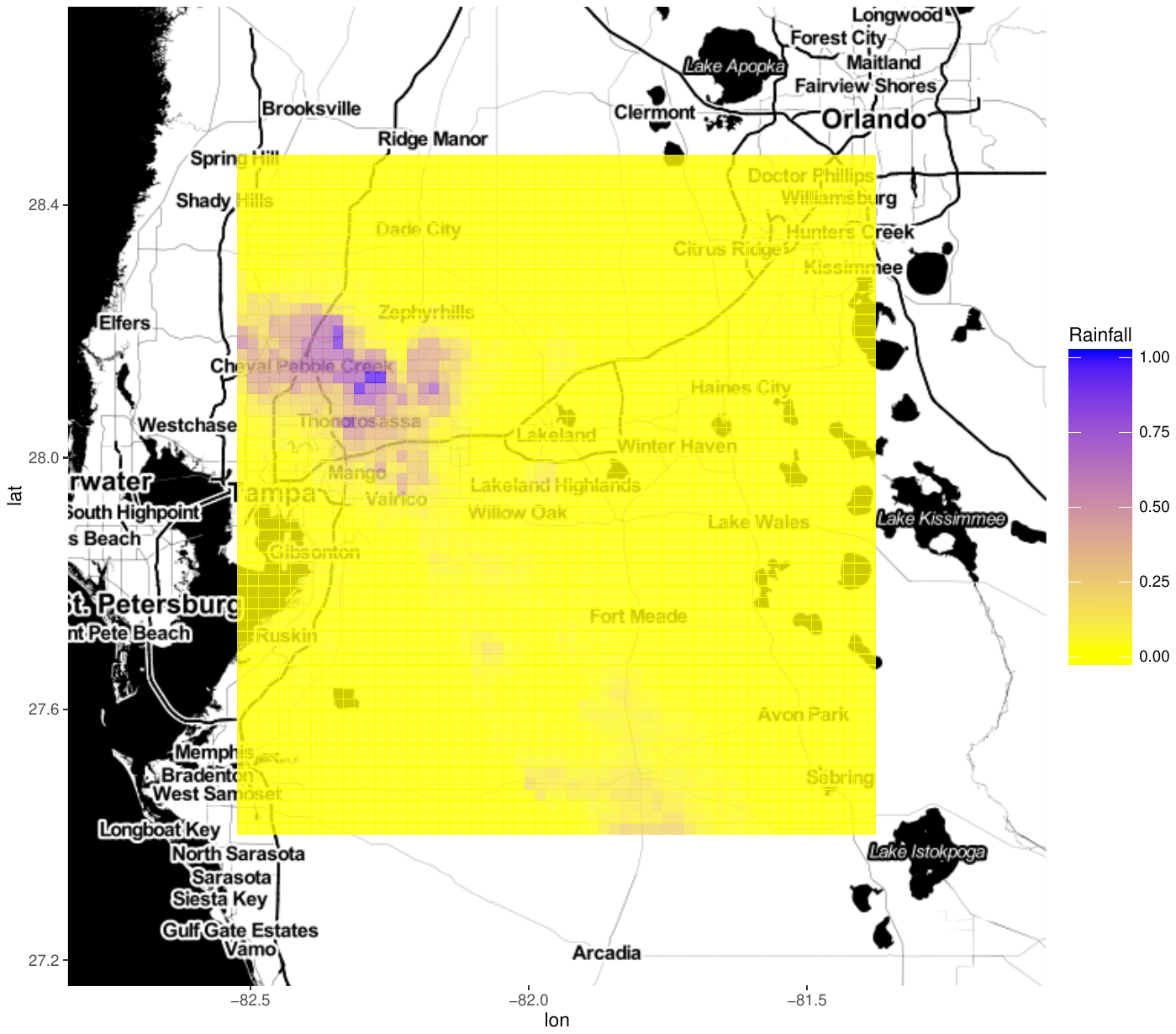} }\\
\rotatebox[origin=c]{90}{Simulations}  &  \raisebox{-.5\height}{\includegraphics[scale = 0.25]{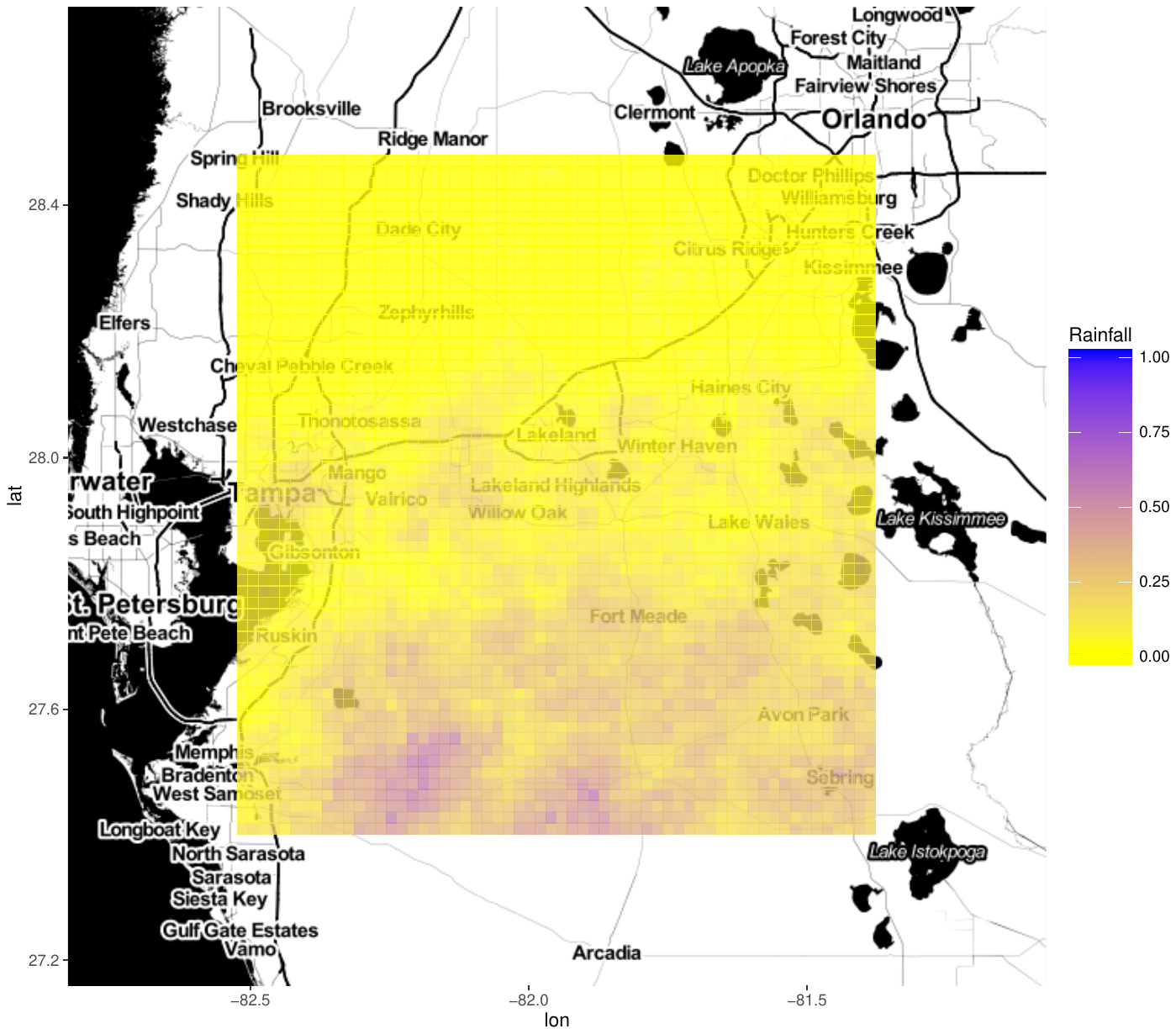}} &   \raisebox{-.5\height}{\includegraphics[scale = 0.25]{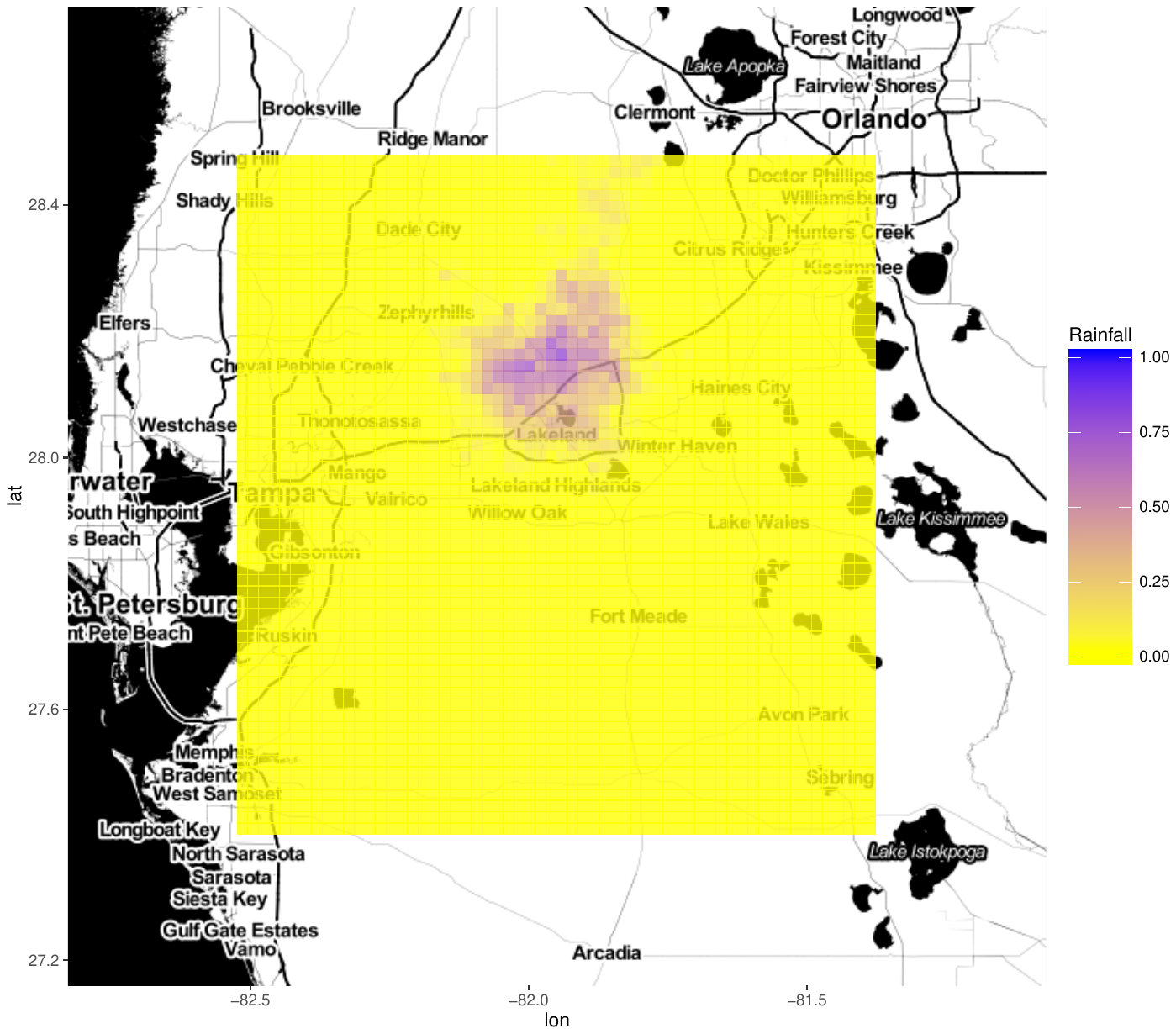}} \\
\end{tabular}
\end{center}
\caption{Fifteen-minute cumulated rainfall (inches), observed (first row) and simulated (second row) for the risk functionals $\R_{\text{sum}}$ (left) and $\R_{\text{max}}$ (right) with an intensity equivalent to top $60$ most intense events.}
\label{fig: rainfall simulations sum}
\end{figure}

Finally, we use the models fitted in Section~\ref{sec: model fitting} to simulate events with intensities equivalent to the $60$ most intense events found by our risk functionals.  Simulation is performed by generating a Pareto process with the fitted dependence structure, as in Section~\ref{sec: simul Pareto}.  Figures~\ref{fig: rainfall simulations sum} shows results for $\R_{\text{sum}}$ and $\R_{\max}$; its upper row contains observations from the database, and the second row shows representative simulations.

The simulations seem reasonable for both risk functionals; they successfully reproduce both the spatial dependence and the intensity of the selected observations.  A closer examination suggests that in both cases  the models produce over-smooth rainfall fields. This could be addressed by improving event selection using risk functionals $\R$ that characterize special spatial structures or physical processes.  Also, as we fail to detect anisotropy, more complex models for dependence that integrate possible stochasticity of the spatial patterns might be worthwhile.

\section{Discussion}

In this paper high-dimensional inference methods for $\R$-Pareto processes associated to log-Gaussian random vectors were developed, implemented and compared.  When simulating from the true model, spectral likelihood estimation performs best, closely followed by gradient score estimation, but censored likelihood estimation was found to perform better with simulations from the domain of attraction, except in cases of weak extremal dependence, where it is outperformed by the gradient score.  Even with computational improvements, use of the censored likelihood is limited to a few hundred variables at most.  The gradient score is a good compromise, attractive for its robustness and because it allows a range of risk functionals while remaining cheap to compute. Empirical work suggests room for improvement of the robustness of the gradient score.

We used these inference methods to study extreme spatial rainfall over Florida. The resulting models can reproduce both spatial patterns and extreme intensity for spatially accumulated and local heavy rainfall. In both cases the fitted model provides a reasonable fit and simulations seem broadly consistent with observations. However, the presence of two very different dependence structures highlights the complexity of extreme rainfall and suggests that a mixture model might be considered. Our model is only a first step towards a spatio-temporal rainfall generator: more complex risk functionals should be considered that take temporal dependence into account.

This paper opens the development of spatio-temporal models for extremes using large climatological datasets, with a view toward a better understanding and estimation of risks associated with natural hazards.

\subsection*{Acknowledgement}
We thank the Swiss National Science Foundation for financial support, and Chin Man Mok for providing the Florida rainfall data, which were supplied by the Southwest Florida Water Management District.

\bibliographystyle{apalike}
\bibliography{library}

\begin{thebibliography}{}

\bibitem[Asadi et~al., 2015]{Asadi2015}
Asadi, P., Davison, A.~C., and Engelke, S. (2015).
\newblock {Extremes on River Networks}.
\newblock {\em Annals of Applied Statisitcs}, 9(4):2023--2050.

\bibitem[Buhl and Kluppelberg, 2016]{Buhl2015}
Buhl, S. and Kluppelberg, C. (2016).
\newblock {Anisotropic Brown--Resnick Space-time Processes: Estimation and
  Model Assessment}.
\newblock {\em Extremes}, 19:627---660.

\bibitem[Castruccio et~al., 2016]{Castruccio2014}
Castruccio, S., Huser, R., and Genton, M.~G. (2016).
\newblock {High-order Composite Likelihood Inference for Max-Stable
  Distributions and Processes}.
\newblock {\em Journal of Computational and Graphical Statistics},
  25:1212--1229.

\bibitem[Chiles and Delfiner, 1999]{Chiles2012a}
Chiles, J.-P. and Delfiner, P. (1999).
\newblock {\em {Geostatistics: Modeling Spatial Uncertainty}}.
\newblock Wiley, New York.

\bibitem[Coles, 2001]{Coles2001}
Coles, S. (2001).
\newblock {\em {An Introduction to Statistical Modeling of Extreme Values}}.
\newblock Springer, London.

\bibitem[Coles and Tawn, 1991]{Coles1991}
Coles, S.~G. and Tawn, J.~A. (1991).
\newblock {Modelling Extreme Multivariate Events}.
\newblock {\em Journal of the Royal Statistical Society, Series B},
  53(2):377--392.

\bibitem[Coles and Tawn, 1996]{Tawn1996}
Coles, S.~G. and Tawn, J.~A. (1996).
\newblock {Modelling Extremes of the Areal Rainfall Process.}
\newblock {\em Journal of the Royal Statistical Society, Series B},
  58(2):329--347.

\bibitem[Davis et~al., 2013]{Davis}
Davis, R.~A., Kluppelberg, C., and Steinkohl, C. (2013).
\newblock {Max-stable Processes for Modelling Extremes Observed in Space and
  Time}.
\newblock {\em Journal of the Korean Statistical Society}, 42(3):399--414.

\bibitem[Davis and Mikosch, 2009]{Davis2009}
Davis, R.~A. and Mikosch, T. (2009).
\newblock {The extremogram: A correlogram for extreme events}.
\newblock {\em Bernoulli}, 15(4):977--1009.

\bibitem[Davison and Smith, 1990]{Davison1990}
Davison, A.~C. and Smith, R.~L. (1990).
\newblock {Models for Exceedances over High Thresholds (with discussion)}.
\newblock {\em Journal of the Royal Statistical Society, Series B},
  52(3):393--442.

\bibitem[Dawid et~al., 2016]{Dawid2014}
Dawid, A.~P., Musio, M., and Ventura, L. (2016).
\newblock {Minimum Scoring Rule Inference}.
\newblock {\em Scandinavian Journal of Statistics}, 43(1):123--138.

\bibitem[de~Haan and Lin, 2001]{DeHaan2001}
de~Haan, L. and Lin, T. (2001).
\newblock {On Convergence Toward an Extreme Value Distribution in $C [0,1]$}.
\newblock {\em The Annals of Probability}, 29(1):467--483.

\bibitem[de~Haan and Resnick, 1993]{DeHaan1993}
de~Haan, L. and Resnick, S.~I. (1993).
\newblock {Estimating the Limit Distribution of Multivariate Extremes}.
\newblock {\em Communications in Statistics. Stochastic Models}, 9(2):275--309.

\bibitem[Dick and Pillichshammer, 2010]{Dick2010}
Dick, J. and Pillichshammer, F. (2010).
\newblock {\em {Digital Nets and Sequences}}.
\newblock Cambridge University Press, Cambridge.

\bibitem[Dombry et~al., 2016]{Dombry2016}
Dombry, C., Engelke, S., and Oesting, M. (2016).
\newblock Exact simulation of max-stable processes.
\newblock {\em Biometrika}, 103:303--317.

\bibitem[Dombry and Ribatet, 2015]{Dombry2013}
Dombry, C. and Ribatet, M. (2015).
\newblock {Functional Regular Variations, Pareto Processes and Peaks Over
  Thresholds}.
\newblock {\em Statistics and Its Interface}, 8(1):9--17.

\bibitem[Einmahl et~al., 2016]{Einmahl2016}
Einmahl, J. H.~J., Kiriliouk, A., Krajina, A., and Segers, J. (2016).
\newblock {An M-estimator of Spatial Tail Dependence}.
\newblock {\em Journal of the Royal Statistical Society, Series B},
  78(1):275--298.

\bibitem[Engelke et~al., 2015]{Engelke2012b}
Engelke, S., Malinowski, A., Kabluchko, Z., and Schlather, M. (2015).
\newblock {Estimation of Huesler--Reiss Distributions and Brown--Resnick
  Processes}.
\newblock {\em Journal of the Royal Statistical Society, Series B},
  77(1):239--265.

\bibitem[Ferreira and de~Haan, 2014]{Ferreira2014}
Ferreira, A. and de~Haan, L. (2014).
\newblock {The generalized Pareto Process; with a View Towards Application and
  Simulation}.
\newblock {\em Bernoulli}, 20(4):1717--1737.

\bibitem[Genz, 2013]{Genz2013}
Genz, A. (2013).
\newblock {QSILATMVNV, Matlab program}.

\bibitem[Genz and Bretz, 2009]{Genz2009}
Genz, A. and Bretz, F. (2009).
\newblock {\em {Computation of Multivariate Normal and {\$}t{\$}
  Probabilities}}.
\newblock Springer, Dordrecht.

\bibitem[Genz et~al., 2014]{Genz2014}
Genz, A., Bretz, F., Miwa, T., Mi, X., Leisch, F., Scheipl, F., and Hothorn, T.
  (2014).
\newblock {mvtnorm: Multivariate Normal and t-Distributions. R package version
  1.0-2.}

\bibitem[Geyer, 1994]{Geyer1994}
Geyer, C.~J. (1994).
\newblock {On the Convergence of Monte Carlo Maximum Likelihood Calculations}.
\newblock {\em Journal of the Royal Statistical Society, Series B},
  56(1):261--274.

\bibitem[Gneiting and Raftery, 2007]{Gneiting2007b}
Gneiting, T. and Raftery, A.~E. (2007).
\newblock {Strictly Proper Scoring Rules, Prediction, and Estimation}.
\newblock {\em Journal of the American Statistical Association},
  102(477):359--378.

\bibitem[Gumbel, 1958]{Gumbel1958}
Gumbel, E.~J. (1958).
\newblock {\em {Statistics of Extremes}}.
\newblock Columbia University Press, New York.

\bibitem[Hult and Lindskog, 2005]{Hult2005}
Hult, H. and Lindskog, F. (2005).
\newblock {Extremal Behavior of Regularly Varying Stochastic Processes}.
\newblock {\em Stochastic Processes and their Applications}, 115(2):249--274.

\bibitem[Huser and Davison, 2013]{Huser2013a}
Huser, R. and Davison, A.~C. (2013).
\newblock {Composite Likelihood Estimation for the Brown--Resnick Process}.
\newblock {\em Biometrika}, 100(2):511--518.

\bibitem[Huser et~al., 2016]{Huser}
Huser, R., Davison, A.~C., and Genton, M.~G. (2016).
\newblock {Likelihood Estimators for Multivariate Extremes}.
\newblock {\em Extremes}, 19(1):79--103.

\bibitem[Hyv{\"{a}}rinen, 2005]{Hyvarinen2005}
Hyv{\"{a}}rinen, A. (2005).
\newblock {Estimation of Non-normalized Statistical Models by Score Matching}.
\newblock {\em Journal of Machine Learning Research}, 6(4):695--708.

\bibitem[Hyv{\"{a}}rinen, 2007]{Hyvarinen2007}
Hyv{\"{a}}rinen, A. (2007).
\newblock {Some Extensions of Score Matching}.
\newblock {\em Computational Statistics {\&} Data Analysis}, 51(5):2499--2512.

\bibitem[Joe et~al., 1992]{Joe.Smith.Weissman:1992}
Joe, H., Smith, R.~L., and Weissman, I. (1992).
\newblock {Bivariate threshold methods for extremes}.
\newblock {\em Journal of the Royal Statistical Society, Series B},
  54:171--183.

\bibitem[Kabluchko et~al., 2009]{Kabluchko2009}
Kabluchko, Z., Schlather, M., and de~Haan, L. (2009).
\newblock {Stationary Max-stable Fields Associated to Negative Definite
  Functions}.
\newblock {\em Annals of Probability}, 37(5):2042--2065.

\bibitem[Kl{\"{u}}ppelberg and Resnick, 2008]{Kluppelberg2008}
Kl{\"{u}}ppelberg, C. and Resnick, S.~I. (2008).
\newblock {The Pareto Copula, Aggregation of Risks, and the Emperor's Socks}.
\newblock {\em Journal of Applied Probability}, 45(1):67--84.

\bibitem[Lindskog et~al., 2014]{Lindskog2014}
Lindskog, F., Resnick, S.~I., and Roy, J. (2014).
\newblock {Regularly Varying Measures on Metric Spaces: Hidden Regular
  Variation and Hidden Jumps}.
\newblock {\em Probability Surveys}, 11:270--314.

\bibitem[Madsen et~al., 1997]{Madsen1997}
Madsen, H., Rasmussen, P.~F., and Rosbjerg, D. (1997).
\newblock {Comparison of Annual Maximum Series and Partial Duration Series
  Methods for Modeling Extreme Hydrologic Events}.
\newblock {\em Water Resources Research}, 33(4):747--757.

\bibitem[Nuyens and Cools, 2004]{Nuyens2006}
Nuyens, D. and Cools, R. (2004).
\newblock {Fast Component-by-Component Construction, a Reprise for Different
  Kernels}.
\newblock In Niederreiter, H. and Talay, D., editors, {\em Monte Carlo and
  Quasi-Monte Carlo Methods 2004}, pages 373--387. Springer Berlin.

\bibitem[Opitz, 2013]{Opitz2013}
Opitz, T. (2013).
\newblock {\em {Extr{\^{e}}mes Multivari{\'{e}}s et Spatiaux : Approches
  Spectrales et Mod{\`{e}}les Elliptiques}}.
\newblock PhD thesis, Universit{\'{e}} Montpellier II.

\bibitem[Padoan et~al., 2010]{Padoan2010}
Padoan, S.~A., Ribatet, M., and Sisson, S.~A. (2010).
\newblock {Likelihood-Based Inference for Max-Stable Processes}.
\newblock {\em Journal of the American Statistical Association},
  105(489):263--277.

\bibitem[Resnick, 2007]{Resnick2007}
Resnick, S.~I. (2007).
\newblock {\em {Heavy-tail Phenomena: Probabilistic and Statistical Modeling}}.
\newblock Springer, New York.

\bibitem[Rootz{\'{e}}n and Tajvidi, 2006]{Rootzen2006}
Rootz{\'{e}}n, H. and Tajvidi, N. (2006).
\newblock {Multivariate Generalized Pareto Distributions}.
\newblock {\em Bernoulli}, 12(5):917--930.

\bibitem[Schlather and Tawn, 2003]{Schlather2003}
Schlather, M. and Tawn, J.~A. (2003).
\newblock {A Dependence Measure for Multivariate and Spatial Extreme Values:
  Properties and Inference}.
\newblock {\em Biometrika}, 90(1):139--156.

\bibitem[Stephenson and Tawn, 2005]{Stephenson2005a}
Stephenson, A.~G. and Tawn, J.~A. (2005).
\newblock {Exploiting Occurrence Times in Likelihood Inference for
  Componentwise Maxima}.
\newblock {\em Biometrika}, 92(1):213--227.

\bibitem[Thibaud et~al., 2016]{Thibaud.etal:2016}
Thibaud, E., Aalto, J., Cooley, D.~S., Davison, A.~C., and Heikkinen, J.
  (2016).
\newblock {Bayesian inference for the Brown--Resnick process, with an
  application to extreme low temperatures}.
\newblock {\em Annals of Applied Statistics}, 10:2303--2324.

\bibitem[Thibaud and Opitz, 2015]{Thibaud2013}
Thibaud, E. and Opitz, T. (2015).
\newblock {Efficient Inference and Simulation for Elliptical Pareto Processes}.
\newblock {\em Biometrika}, 102(4):855--870.

\bibitem[Thorarinsdottir et~al., 2013]{Thorarinsdottir2013}
Thorarinsdottir, T.~L., Gneiting, T., and Gissibl, N. (2013).
\newblock {Using Proper Divergence Functions to Evaluate Climate Models}.
\newblock {\em SIAM/ASA Journal on Uncertainty Quantification}, 1(1):522--534.

\bibitem[Varin et~al., 2011]{Varin.Reid.Firth:2011}
Varin, C., Reid, N., and Firth, D. (2011).
\newblock {An overview of composite marginal likelihoods}.
\newblock {\em {Statistica Sinica}}, 21:5--42.

\bibitem[Wadsworth, 2015]{Wadsworth2015}
Wadsworth, J.~L. (2015).
\newblock {On the Occurrence Times of Componentwise Maxima and Bias in
  Likelihood Inference for Multivariate Max-stable Distributions}.
\newblock {\em Biometrika}, 102(3):705--711.

\bibitem[Wadsworth and Tawn, 2014]{Wadsworth2014}
Wadsworth, J.~L. and Tawn, J.~A. (2014).
\newblock {Efficient Inference for Spatial Extreme Value Processes Associated
  to Log-Gaussian Random Functions}.
\newblock {\em Biometrika}, 101(1):1--15.

\end{thebibliography}

\appendix
\section{High-dimensional censored likelihood}\label{app: censored likelihood}
\subsection{Computational considerations}\label{app: parallel censored likelihood}

The algorithm due to  \citet{Genz2009} and implemented in the R package \verb+mvtnorm+   \citep{Genz2014} provides an unbiased estimate of a multivariate normal probabilities, with an indication of its largest probable error. An improved Matlab implementation \citep{Genz2013} makes better use of quasi-Monte Carlo methods. We translated this code into \verb C++  to speed it up; see Appendix~\ref{app: mvn estim}.

Function evaluation is independent for each sample, so we also adapted the algorithm for GPU computing and compared different implementations.
Our \verb C++  implementation is about $4$ times faster than the \verb mvtnorm  implementation for a probable worst-case error of order $10^{-3}$.
GPU computing provides a slight improvement in speed compared to C++ for reasonably low error, but shows a significant speed-up for higher accuracies ($\lesssim 10^{-4}$).
A computation time of 1~s for estimation of one integral seems reasonable for censored likelihood, and is achievable for $I\approx 500$ for probable worst-case errors of order $10^{-3}$ without GPU computing.

Although Jensen's inequality implies that estimation of the log-likelihood function is biased for finite $p$, quasi-Monte Carlo estimation of an integral is unbiased, so for a sufficiently high $p$,
\begin{equation}\label{eq: estim approx}
\log \Phi^p = \log (\Phi + \epsilon^p) = \log (\Phi) + \frac{\epsilon^p}{\Phi} + o_p\left(\frac{\epsilon^p}{\Phi}\right), 
\end{equation}
where $\epsilon^p$ is a random error with zero mean and bounded variance.
Using equation~(\ref{eq: estim approx}) with a small $\epsilon^p$, we have $\widehat{\theta}_{\text{cens}} \approx \Es\left(\widehat{\theta}^p_{\text{cens}}\right)$.
On a multi-node cluster, for scalability purposes, it is more efficient to combine independent estimates $\widehat{\theta}^p_{\text{cens},q}$ ($q = 1,\dots, \bar{p}$) into $\widetilde{\theta}^{\bar{p}} = \bar{p}^{-1} \sum_{i = q}^{\bar{p}} \widehat{\theta}^p_{\text{cens},q}$  than to compute a single estimate $\widehat{\theta}^{p \times \bar{p}}_{\text{cens}}$ with $p \times \bar{p}$ samples in the quasi-Monte Carlo procedures.
Indeed, maximization of $\ell^p_{\text{cens}}(\theta)$ requires a reduction step, in which the computations performed on each node are assembled, for every evaluation of the objective function. 
Hence for a cluster with several nodes, where communication is usually slow and reduction steps expensive, $\widetilde{\theta}^{\bar{p}} $  is more efficient because the computation of several $\widehat{\theta}^p_{\text{cens},q}$ can be done independently on different nodes.
Moreover, use of $\widetilde{\theta}^{\overline{p}}$ allows $\text{var}(\widehat{\theta}^p_{\text{cens}})$ to be estimated.

We parallelized the above inference procedure on a cluster with $12$ nodes each of $16$ cores.
First computation of $\ell^p_{\text{cens}}(\theta)$ was parallelized within each node using the R package \verb+parallel+.  
The time needed to compute the censored likelihood for a $300$-dimensional vector for a generalized Pareto process associated to a log-Gaussian random function with $p = 499$ and different dependence strengths dropped from minutes to a dozen seconds.
Each node performs an independent maximization using the R routine \verb+optim+. Even if slightly biased, this approach is computationally efficient for our cluster infrastructure.
If the empirical variance of $\widehat{\theta}^p_{\text{cens}}$ is too high then the number of samples $p$ should be increased.
For high accuracy and/or complex models, GPU computing may be relevant.
Lastly, the tolerance of the optimization algorithm must be reduced for low $p$ to ensure its convergence if the quasi-Monte Carlo estimates vary substantially.

\subsection{Algorithm for multivariate normal distribution function estimation} \label{app: mvn estim}
This algorithm is a simplified version of that of \citet{Genz2009}. To estimate the $I$-dimensional multivariate normal distribution $\Phi_I(x, \Sigma)$:
\begin{enumerate}
  \item input covariance matrix $\Sigma$, upper bound $x$, number of deterministic samples $p$, number of random shifts $p'$ and generating vector $v$;
  \item compute lower triangular Cholesky factor $\textbf{L}$ for $\Sigma$, permuting $x$, and rows and columns of $\Sigma$ for variable prioritisation;
  \item initialize $\Phi = 0$, $\delta = 0$ and $V = 0$;
  \item for $q'$  in $1,\ldots,p'$:
  \begin{enumerate}
  \item set $I_q' = 0$ and generate uniform random shift $\Delta \in [0,1]^I$;
  \item for $q$ in $1,\ldots, p$:
  \begin{enumerate}[(i)]
  \item set $z_q = |2 \times  \overline{( qv + \Delta)} - 1|$
  
  $e_1 = \Phi(b_1/l_{1,1})$
  
  $f_1 = e_1$;
  \item for $i$ in $2,\ldots, I$
  
  set $y_{i-1} = \Phi^{-1}(w_{i-1} e_{i-1})$
  
  $e_i = \Phi\left(\frac{b_i -\sum_{j = 1}^{i-1}l_{i,j}y_j}{l_{i,i}}\right)$
  
  $f_i = e_i f_{i-1}$
  
  End $i$ loop;
  \item set $I_{q'} = I_ {q'}+ (f_i-I_{q'})/q$;
\end{enumerate}
  End $q$ loop;
  
  \item Set $\delta = (I_{q'} - \Phi)/i$, $\Phi = \Phi + \delta$, $V = (q' -2)V/i + \delta^2$ and ${\rm ERR} = \alpha\sqrt{V}$;
\end{enumerate}
end $q'$ loop;
\item output $\Phi \approx \Phi_k(-\infty, x;\Sigma)$ with error estimate ${\rm ERR}$.
\end{enumerate}

\section{Gradient score for Brown--Resnick processes} \label{app: BR score}
\cite{Wadsworth2014} derive an alternative expression for the intensity function (\ref{eq: cdfBR Engelke}):
\begin{equation}\label{eq: cdfBR Wadsworth}
\begin{aligned}
\lambda_{\theta}(x) = & \frac{|\det \Sigma_{\theta}^*|^{-1/2}(1^T_I \rho)^{-1/2}}{(2\pi)^{(I-1)/2} x_1 \cdots x_I} \exp \left(-\frac{1}{2} \left[ \log x^T \Gamma \log x + \log x^T \left\{ \frac{2\rho}{1^T_I \rho}  +  \left(\Sigma^*_{\theta}\right)^{-1} \sigma - \frac{\rho\rho^T\sigma}{1^T_I \rho}\right\}\right]\right) \\
& \times \exp \left[-\frac{1}{2} \left\{\frac{1}{4} \sigma^T\left(\Sigma^*_{\theta}\right)^{-1}\sigma  - \frac{1}{4} \frac{\sigma^T \rho\rho^T \sigma}{1^T_I \rho} + \frac{\sigma^T \rho}{1^T_I \rho} - \frac{1}{1^T_I \rho}\right\} \right] , \quad x \in A_{\R}(u),
\end{aligned}
\end{equation}
where $\Sigma^*_\theta$ is the $I$-dimensional covariance matrix of a non-stationary Gaussian process with semi-variogram $\gamma$, $\rho = \left(\Sigma^*_{\theta}\right)^{-1}1_I$, $\Gamma = \left(\Sigma^*_{\theta}\right)^{-1} - \rho\rho^T/1^T_I \rho$ and $\sigma = \text{diag}(\Sigma^*_{\theta})$.
This expression is symmetric and thus it is more convenient to compute its gradient and Laplacian.

The gradient of the density function $\lambda_{\theta,u}^{\R}$ with respect to $x$ and with the notation of equation (\ref{eq: cdfBR Wadsworth}) is
\begin{equation}
\nabla_{x} \log  \lambda_{\theta,u}^{\R} (x) = 	- \Gamma \log x \otimes \frac{1}{x} - \frac{1}{2x} \otimes \left(\frac{2\rho}{1^T_I \rho} + 2 + \Gamma^{-1}\sigma - \frac{\rho\rho^T \sigma}{1^T_I \rho}\right), \quad x \in A_{\R}(u), \quad u >0,
\end{equation}
where $\otimes$ is the Hadamard product, $1_I$ is a $I$-dimensional vector with unit components, $\Sigma^*_\theta$ is the covariance matrix of the non-stationary Gaussian process with semi-variogram $\gamma_\theta$, $\rho = \left(\Sigma^*_\theta\right)^{-1}1_I$, $\Gamma = \left(\Sigma^*_\theta\right)^{-1} - \rho\rho^T/1^T_I\rho$ and $\sigma = \text{diag}(\Sigma_\theta^*)$.
The Laplacian of this density function, $\triangle_{x} \log  \lambda_{\theta,u}^{\R}(x)$, equals
\begin{equation}
-\text{diag}(\Gamma)^T\left[ \frac{1-\log x}{x^2}\right] + \left\|(\Gamma - \text{diag}(\Gamma)) \log x \otimes \frac{1}{x^2}\right\|_1  + \frac{1}{(2x^2)^T} \left\{\frac{2\rho}{1^T_I \rho} + 2 + \left(\Sigma^*_\theta\right)^{-1}\sigma - \frac{\rho\rho^T \sigma}{1^T_I \rho}\right\},
\end{equation}
where $x \in A_{\R}(u)$, $u >0$ and $\| \cdot \|_1$ denotes the $L_1$ norm.

\section{Average computation times of the fitting procedures} \label{app: computation time}

\begin{table}[H]
\begin{center}
\begin{tabular}{ c  c  c  c  c }
 Grid size & $\kappa$ & Spectral likelihood & Censored log-likelihood & Gradient score \\ \hline
\multirow{3}{*}{$10 \times 10$} & 0.5 &4  & 135 & 6 \\
& 1 &4  & 140 & 4.9 \\ 
& 1.3 &4.5  & 129 & 4.8 \\ 
\\
\multirow{3}{*}{$20 \times 10$} & 0.5 &14.3  &486& 10 \\ 
& 1 &6  & 492 & 9.7 \\ 
& 1.3 &6.7  & 483 & 9.8 \\ 
\\
\multirow{3}{*}{$20 \times 15$} & 0.5 &14  & 1190 & 18 \\ 
& 1 &14  & 1217 & 16.4 \\ 
& 1.3 &14.6  & 1236 & 18.8 \\ 
\end{tabular}
\end{center}
\caption{Average times (s) of the optimization for the different objective functions, when fitting a Brown--Resnick process applied to the three different semi-variogram models $\gamma$ with $\kappa= \{0.5,1,1.3\}$ and the three grids  $10 \times 10$, $20 \times 10$ and $20 \times 15$. Random starting points are used for fair comparison.} 
\label{tab: time est Pareto}
\end{table}

\section{Proof of the Proposition} \label{app: mda score normality}
Let $(\mathrm{y}^m)_{m = 1, \dots, n}$ be independent replicates of a regularly-varying random vector $Y$ with normalized marginals and measure $\nu_{\theta_0}$.
Let $k_u = k_u(n)$ be a sequence of integers, where $n$ is the sample size, statisfying $ k_u(n) \rightarrow \infty$ and $ k_u(n) = o(n)$ as $n \rightarrow \infty$ and suppose we only keep vectors such that $\left\{\R(\mathrm{y}^m)\right\}_{m = 1, \dots, n}$ exceeds the threshold $n/k_u$, i.e., we retain the set 
$$
A_{\R}\left(\frac{n}{k_u}\right) = \left\{ \tilde{\mathrm{y}} : \R\left(\tilde{\mathrm{y}}\right) = \R\left(\frac{k_u}{n}\mathrm{y}\right) > 1 \right\}.
$$
For any $A \in \mathbb{R}^I_+$, we first need the asymptotic normality of the empirical measure
$$
\tilde{\nu}_{k_u}\left(A\right) = \frac{1}{k_u} \sum_{m = 1}^n \mathbb{1}\left(\tilde{\mathrm{y}}^m \in A \right)
$$
Since $G$ is in the max--domain of attraction of $P$, Proposition~2.1 in \citet{DeHaan1993} gives
\begin{equation}\label{eq: convergence of measure}
\tilde{\nu}_{k_u}\left(A\right) \xrightarrow{\text{Pr}} \nu\left(A\right), \quad A \in \mathbb{R}^I_+, \quad n \rightarrow \infty,
\end{equation}
where $\nu$ is the exponent measure associated to the multivariate extreme value distribution $P$ and $\xrightarrow{\text{Pr}}$ denotes convergence in probability.
Moreover, following Propositions~3.1 and~3.2 in \citet{DeHaan1993}, define the random field $Z_n(x)$, $x \in (0, \infty]^I$, by
$$
Z_n(x) = \sqrt{k_u} \left\{  \tilde{\nu}_{k_u}\left((0, x]\right) -  \tilde{\nu}\left((0, x]^c\right) \right\} , \quad  x \in (0, \infty]^I.
$$
There exists a zero-mean Gaussian random field $Z(x)$, $x \in (0, \infty]^I$, with continuous sample paths and covariance function
$$
\text{Cov}\left\{Z_n(x^1), Z_n(x^2) \right\} = \nu \left\{ (0, x^1]^c \cap (0, x^2]^c \right\}, \quad x^1,x^2 \in  (0, \infty]^I,
$$
such that $Z_n(x) $ converges weakly to $Z(x)$ in the space of cadlag functions defined on $ (0, \infty]^I$ equipped with the Skorohod topology.

Now let $\delta$ be a proper scoring rule satisfying the regularity conditions of Theorem 4.1~of \citet{Dawid2014}.
The maximum scoring rule estimator $\widehat{\theta}_{k_u}^\delta$ is defined by
$$
\sum_{\left\{m, \mathrm{y}^m \in A_{\R}\left(n/k_u\right)\right\}} \nabla_\theta \delta\left(\widehat{\theta}_{\delta,k_u}^{\R}, \mathrm{y}^m \right) = 0,
$$
which is equivalent to
$$
\frac{1}{k_u} \int_{A_{\R}\left(n/k_u\right)} \nabla_\theta \delta\left(\widehat{\theta}_{\delta,k_u}^{\R}, \mathrm{y} \right) \tilde{\nu}_{k_u}\left(\text{d}\mathrm{y}\right)= 0.
$$
The second-order condition in the hypothesis of Theorem 4.1 in \citet{Dawid2014} allows us to use a Taylor expansion around $\theta_0$, yielding
$$
\frac{1}{k_u}\int_{A_{\R}\left(n/k_u\right)} \nabla_\theta \delta\left(\theta_0, \mathrm{y} \right) \tilde{\nu}_{k_u}\left(\text{d}\mathrm{y}\right) + \left(\widehat{\theta}_{\delta,k_u}^{\R} - \theta_0\right) \frac{1}{k_u}\int_{A_{\R}\left(n/k_u\right)} \nabla_\theta^2 \delta\left(\theta_0, \mathrm{y} \right) \tilde{\nu}_{k_u}\left(\text{d}\mathrm{y}\right) + o\left\{\left(\widehat{\theta}_{\delta,k_u}^{\R} - \theta_0\right)\right\} = 0.
$$
Also equation~(\ref{eq: convergence of measure}) ensures that
$$
\frac{1}{k_u}\int_{A_{\R}\left(n/k_u\right)} \nabla_\theta^2 \delta\left(\theta_0, \mathrm{y} \right) \tilde{\nu}_{k_u}\left(\text{d}\mathrm{y}\right) \xrightarrow{\text{Pr}} \text{E}_P\left\{\frac{\partial^2\delta}{\partial \theta^2}(\theta_0) \right\} = K,
$$
and using the convergence of $Z_n$, we get
$$
\frac{1}{k_u}\int_{A_{\R}\left(n/k_u\right)} \nabla_\theta \delta\left(\theta_0, \mathrm{y} \right) \tilde{\nu}_{k_u}\left(\text{d}\mathrm{y}\right)  \xrightarrow{D} \mathcal{N}\left[0, \Es_P \left\{\frac{\partial\delta}{\partial \theta}(\theta_0)\frac{\partial\delta}{\partial \theta}(\theta_0)^T\right\}\right], \quad n \rightarrow \infty.
$$
Then it is straightforward to see that
$$
\sqrt{n_u}\left(\widehat{\theta}_{\delta,k_u}^{\R} - \theta_0\right) \xrightarrow{D} \mathcal{N}\left\{0, K^{-1}J\left(K^{-1}\right)^T\right\}, \quad n \rightarrow \infty,
$$
with $J = \Es_P \left\{\partial\delta/\partial \theta(\theta_0)\partial\delta/\partial \theta(\theta_0)^T\right\}$.

\section{Pareto process simulation}\label{app: pareto sims}
To compare the performance of our estimators in Section \ref{sec: simul Pareto}, the simulation of a Pareto $P$ process for $I > 0 $ locations over $[0,100]^2$ with semi-variogram $\gamma$ is done as follows:
\begin{itemize}
  \item for regularly spaced locations $\{s_1,\ldots, s_I\}\in [0,100]^2$, choose $i\in\{1,\ldots, I\}$ uniformly at random;
  \item for a given semi-variogram $\gamma(s,s')$, $s,s' \in [0,100]^2$, generate an $(I-1)$-dimensional Gaussian vector $Z$ with covariance matrix $\Sigma  = \{ \gamma(s_j,s_i) + \gamma(s_k,s_i) - \gamma(s_j,s_k) \}_{ j,k \in \left\{1, \dots, I\right\} \setminus \{i\}}$ and mean $\mu = \{- \gamma(s_j,s_j)\}_{ j \in \left\{1, \dots, I\right\} \setminus \{i\}}$, i.e., conditional on the value at $s_i$; 
  \item  set $Q_i = 1$ and $Q_1=\exp(Z_1), \dots, Q_{i-1}=\exp(Z_{i-1}), Q_{i+1}=\exp(Z_i),\dots, Q_{I} = \exp( Z_{I-1})$;
   \item generate  a Pareto random variable $U$ with distribution function $1 - 1/x$ ($x > 1$) and set $P = U Q/ \|Q\|_1$;
   \item return $P$.
\end{itemize}

\end{document}